\documentclass[longauth,traditabstract]{aa} 
\usepackage{amsmath}


\def\setsymbol#1#2{\expandafter\def\csname #1\endcsname{#2}}
\def\getsymbol#1{\csname #1\endcsname}

\def\Planck{\textit{Planck}}

\def\all2013resultspapers{\nocite{planck2013-p01, planck2013-p02, planck2013-p02a, planck2013-p02d, planck2013-p02b, planck2013-p03, planck2013-p03c, planck2013-p03f, planck2013-p03d, planck2013-p03e, planck2013-p01a, planck2013-p06, planck2013-p03a, planck2013-pip88, planck2013-p08, planck2013-p11, planck2013-p12, planck2013-p13, planck2013-p14, planck2013-p15, planck2013-p05b, planck2013-p17, planck2013-p09, planck2013-p09a, planck2013-p20, planck2013-p19, planck2013-pipaberration, planck2013-p05, planck2013-p05a, planck2013-pip56, planck2013-p06b}}

\newbox\tablebox    \newdimen\tablewidth
\def\leaderfil{\leaders\hbox to 5pt{\hss.\hss}\hfil}
\def\tablenote#1 #2\par{\begingroup \parindent=0.8em
    \abovedisplayshortskip=0pt\belowdisplayshortskip=0pt
    \noindent
    $$\hss\vbox{\hsize\tablewidth \hangindent=\parindent \hangafter=1 \noindent
    \hbox to \parindent{$^#1$\hss}\strut#2\strut\par}\hss$$
    \endgroup}

\def\L2{\ifmmode L_2\else $L_2$\fi}

\def\DeltaT{\ifmmode \Delta T\else $\Delta T$\fi}
\def\deltat{\ifmmode \Delta t\else $\Delta t$\fi}
\def\fknee{\ifmmode f_{\rm knee}\else $f_{\rm knee}$\fi}
\def\Fmax{\ifmmode F_{\rm max}\else $F_{\rm max}$\fi}
\def\lsim{\mathrel{\raise .4ex\hbox{\rlap{$<$}\lower 1.2ex\hbox{$\sim$}}}}
\def\gsim{\mathrel{\raise .4ex\hbox{\rlap{$>$}\lower 1.2ex\hbox{$\sim$}}}}
\def\ra[#1 #2 #3.#4]{#1\sup{h}#2\sup{m}#3\sup{s}\llap.#4}
\def\dec[#1 #2 #3.#4]{#1\deg#2\arcm#3\arcs\llap.#4}
\def\deco[#1 #2 #3]{#1\deg#2\arcm#3\arcs}
\def\rra[#1 #2]{#1\sup{h}#2\sup{m}}

\def\dots{\relax\ifmmode \ldots\else $\ldots$\fi}
\def\WHzsr{\ifmmode $W\,Hz\mo\,sr\mo$\else W\,Hz\mo\,sr\mo\fi}
\def\mHz{\ifmmode $\,mHz$\else \,mHz\fi}
\def\GHz{\ifmmode $\,GHz$\else \,GHz\fi}
\def\mKs{\ifmmode $\,mK\,s$^{1/2}\else \,mK\,s$^{1/2}$\fi}
\def\muKs{\ifmmode \,\mu$K\,s$^{1/2}\else \,$\mu$K\,s$^{1/2}$\fi}
\def\muKRJs{\ifmmode \,\mu$K$_{\rm RJ}$\,s$^{1/2}\else \,$\mu$K$_{\rm RJ}$\,s$^{1/2}$\fi}
\def\muKHz{\ifmmode \,\mu$K\,Hz$^{-1/2}\else \,$\mu$K\,Hz$^{-1/2}$\fi}
\def\MJysr{\ifmmode \,$MJy\,sr\mo$\else \,MJy\,sr\mo\fi}
\def\MJysrmK{\ifmmode \,$MJy\,sr\mo$\,mK$_{\rm CMB}\mo\else \,MJy\,sr\mo\,mK$_{\rm CMB}\mo$\fi}
\def\microns{\ifmmode \,\mu$m$\else \,$\mu$m\fi}

\def\muK{\ifmmode \,\mu$K$\else \,$\mu$\hbox{K}\fi}
\def\microK{\ifmmode \,\mu$K$\else \,$\mu$\hbox{K}\fi}
\def\muW{\ifmmode \,\mu$W$\else \,$\mu$\hbox{W}\fi}
\def\kms{\ifmmode $\,km\,s$^{-1}\else \,km\,s$^{-1}$\fi}
\def\kmsMpc{\ifmmode $\,\kms\,Mpc\mo$\else \,\kms\,Mpc\mo\fi}
\setsymbol{LFI:center:frequency:70GHz:units}{70.3\,GHz}
\setsymbol{LFI:center:frequency:44GHz:units}{44.1\,GHz}
\setsymbol{LFI:center:frequency:30GHz:units}{28.5\,GHz}

\setsymbol{LFI:center:frequency:70GHz}{70.3}
\setsymbol{LFI:center:frequency:44GHz}{44.1}
\setsymbol{LFI:center:frequency:30GHz}{28.5}

\setsymbol{LFI:center:frequency:LFI18:Rad:M:units}{71.7\GHz}
\setsymbol{LFI:center:frequency:LFI19:Rad:M:units}{67.5\GHz}
\setsymbol{LFI:center:frequency:LFI20:Rad:M:units}{69.2\GHz}
\setsymbol{LFI:center:frequency:LFI21:Rad:M:units}{70.4\GHz}
\setsymbol{LFI:center:frequency:LFI22:Rad:M:units}{71.5\GHz}
\setsymbol{LFI:center:frequency:LFI23:Rad:M:units}{70.8\GHz}
\setsymbol{LFI:center:frequency:LFI24:Rad:M:units}{44.4\GHz}
\setsymbol{LFI:center:frequency:LFI25:Rad:M:units}{44.0\GHz}
\setsymbol{LFI:center:frequency:LFI26:Rad:M:units}{43.9\GHz}
\setsymbol{LFI:center:frequency:LFI27:Rad:M:units}{28.3\GHz}
\setsymbol{LFI:center:frequency:LFI28:Rad:M:units}{28.8\GHz}
\setsymbol{LFI:center:frequency:LFI18:Rad:S:units}{70.1\GHz}
\setsymbol{LFI:center:frequency:LFI19:Rad:S:units}{69.6\GHz}
\setsymbol{LFI:center:frequency:LFI20:Rad:S:units}{69.5\GHz}
\setsymbol{LFI:center:frequency:LFI21:Rad:S:units}{69.5\GHz}
\setsymbol{LFI:center:frequency:LFI22:Rad:S:units}{72.8\GHz}
\setsymbol{LFI:center:frequency:LFI23:Rad:S:units}{71.3\GHz}
\setsymbol{LFI:center:frequency:LFI24:Rad:S:units}{44.1\GHz}
\setsymbol{LFI:center:frequency:LFI25:Rad:S:units}{44.1\GHz}
\setsymbol{LFI:center:frequency:LFI26:Rad:S:units}{44.1\GHz}
\setsymbol{LFI:center:frequency:LFI27:Rad:S:units}{28.5\GHz}
\setsymbol{LFI:center:frequency:LFI28:Rad:S:units}{28.2\GHz}

\setsymbol{LFI:center:frequency:LFI18:Rad:M}{71.7}
\setsymbol{LFI:center:frequency:LFI19:Rad:M}{67.5}
\setsymbol{LFI:center:frequency:LFI20:Rad:M}{69.2}
\setsymbol{LFI:center:frequency:LFI21:Rad:M}{70.4}
\setsymbol{LFI:center:frequency:LFI22:Rad:M}{71.5}
\setsymbol{LFI:center:frequency:LFI23:Rad:M}{70.8}
\setsymbol{LFI:center:frequency:LFI24:Rad:M}{44.4}
\setsymbol{LFI:center:frequency:LFI25:Rad:M}{44.0}
\setsymbol{LFI:center:frequency:LFI26:Rad:M}{43.9}
\setsymbol{LFI:center:frequency:LFI27:Rad:M}{28.3}
\setsymbol{LFI:center:frequency:LFI28:Rad:M}{28.8}
\setsymbol{LFI:center:frequency:LFI18:Rad:S}{70.1}
\setsymbol{LFI:center:frequency:LFI19:Rad:S}{69.6}
\setsymbol{LFI:center:frequency:LFI20:Rad:S}{69.5}
\setsymbol{LFI:center:frequency:LFI21:Rad:S}{69.5}
\setsymbol{LFI:center:frequency:LFI22:Rad:S}{72.8}
\setsymbol{LFI:center:frequency:LFI23:Rad:S}{71.3}
\setsymbol{LFI:center:frequency:LFI24:Rad:S}{44.1}
\setsymbol{LFI:center:frequency:LFI25:Rad:S}{44.1}
\setsymbol{LFI:center:frequency:LFI26:Rad:S}{44.1}
\setsymbol{LFI:center:frequency:LFI27:Rad:S}{28.5}
\setsymbol{LFI:center:frequency:LFI28:Rad:S}{28.2}

\setsymbol{LFI:white:noise:sensitivity:70GHz:units}{134.7\muKs}
\setsymbol{LFI:white:noise:sensitivity:44GHz:units}{164.7\muKs}
\setsymbol{LFI:white:noise:sensitivity:30GHz:units}{143.4\muKs}

\setsymbol{LFI:white:noise:sensitivity:70GHz}{134.7}
\setsymbol{LFI:white:noise:sensitivity:44GHz}{164.7}
\setsymbol{LFI:white:noise:sensitivity:30GHz}{143.4}

\setsymbol{LFI:white:noise:sensitivity:LFI18:Rad:M:units}{512.0\muKs}
\setsymbol{LFI:white:noise:sensitivity:LFI19:Rad:M:units}{581.4\muKs}
\setsymbol{LFI:white:noise:sensitivity:LFI20:Rad:M:units}{590.8\muKs}
\setsymbol{LFI:white:noise:sensitivity:LFI21:Rad:M:units}{455.2\muKs}
\setsymbol{LFI:white:noise:sensitivity:LFI22:Rad:M:units}{492.0\muKs}
\setsymbol{LFI:white:noise:sensitivity:LFI23:Rad:M:units}{507.7\muKs}
\setsymbol{LFI:white:noise:sensitivity:LFI24:Rad:M:units}{462.2\muKs}
\setsymbol{LFI:white:noise:sensitivity:LFI25:Rad:M:units}{413.6\muKs}
\setsymbol{LFI:white:noise:sensitivity:LFI26:Rad:M:units}{478.6\muKs}
\setsymbol{LFI:white:noise:sensitivity:LFI27:Rad:M:units}{277.7\muKs}
\setsymbol{LFI:white:noise:sensitivity:LFI28:Rad:M:units}{312.3\muKs}
\setsymbol{LFI:white:noise:sensitivity:LFI18:Rad:S:units}{465.7\muKs}
\setsymbol{LFI:white:noise:sensitivity:LFI19:Rad:S:units}{555.6\muKs}
\setsymbol{LFI:white:noise:sensitivity:LFI20:Rad:S:units}{623.2\muKs}
\setsymbol{LFI:white:noise:sensitivity:LFI21:Rad:S:units}{564.1\muKs}
\setsymbol{LFI:white:noise:sensitivity:LFI22:Rad:S:units}{534.4\muKs}
\setsymbol{LFI:white:noise:sensitivity:LFI23:Rad:S:units}{542.4\muKs}
\setsymbol{LFI:white:noise:sensitivity:LFI24:Rad:S:units}{399.2\muKs}
\setsymbol{LFI:white:noise:sensitivity:LFI25:Rad:S:units}{392.6\muKs}
\setsymbol{LFI:white:noise:sensitivity:LFI26:Rad:S:units}{418.6\muKs}
\setsymbol{LFI:white:noise:sensitivity:LFI27:Rad:S:units}{302.9\muKs}
\setsymbol{LFI:white:noise:sensitivity:LFI28:Rad:S:units}{285.3\muKs}

\setsymbol{LFI:white:noise:sensitivity:LFI18:Rad:M}{512.0}
\setsymbol{LFI:white:noise:sensitivity:LFI19:Rad:M}{581.4}
\setsymbol{LFI:white:noise:sensitivity:LFI20:Rad:M}{590.8}
\setsymbol{LFI:white:noise:sensitivity:LFI21:Rad:M}{455.2}
\setsymbol{LFI:white:noise:sensitivity:LFI22:Rad:M}{492.0}
\setsymbol{LFI:white:noise:sensitivity:LFI23:Rad:M}{507.7}
\setsymbol{LFI:white:noise:sensitivity:LFI24:Rad:M}{462.2}
\setsymbol{LFI:white:noise:sensitivity:LFI25:Rad:M}{413.6}
\setsymbol{LFI:white:noise:sensitivity:LFI26:Rad:M}{478.6}
\setsymbol{LFI:white:noise:sensitivity:LFI27:Rad:M}{277.7}
\setsymbol{LFI:white:noise:sensitivity:LFI28:Rad:M}{312.3}
\setsymbol{LFI:white:noise:sensitivity:LFI18:Rad:S}{465.7}
\setsymbol{LFI:white:noise:sensitivity:LFI19:Rad:S}{555.6}
\setsymbol{LFI:white:noise:sensitivity:LFI20:Rad:S}{623.2}
\setsymbol{LFI:white:noise:sensitivity:LFI21:Rad:S}{564.1}
\setsymbol{LFI:white:noise:sensitivity:LFI22:Rad:S}{534.4}
\setsymbol{LFI:white:noise:sensitivity:LFI23:Rad:S}{542.4}
\setsymbol{LFI:white:noise:sensitivity:LFI24:Rad:S}{399.2}
\setsymbol{LFI:white:noise:sensitivity:LFI25:Rad:S}{392.6}
\setsymbol{LFI:white:noise:sensitivity:LFI26:Rad:S}{418.6}
\setsymbol{LFI:white:noise:sensitivity:LFI27:Rad:S}{302.9}
\setsymbol{LFI:white:noise:sensitivity:LFI28:Rad:S}{285.3}

\setsymbol{LFI:knee:frequency:70GHz:units}{29.5\mHz}
\setsymbol{LFI:knee:frequency:44GHz:units}{56.2\mHz}
\setsymbol{LFI:knee:frequency:30GHz:units}{113.7\mHz}

\setsymbol{LFI:knee:frequency:70GHz}{29.5}
\setsymbol{LFI:knee:frequency:44GHz}{56.2}
\setsymbol{LFI:knee:frequency:30GHz}{113.7}

\setsymbol{LFI:knee:frequency:LFI18:Rad:M:units}{16.3\mHz}
\setsymbol{LFI:knee:frequency:LFI19:Rad:M:units}{15.1\mHz}
\setsymbol{LFI:knee:frequency:LFI20:Rad:M:units}{18.7\mHz}
\setsymbol{LFI:knee:frequency:LFI21:Rad:M:units}{37.2\mHz}
\setsymbol{LFI:knee:frequency:LFI22:Rad:M:units}{12.7\mHz}
\setsymbol{LFI:knee:frequency:LFI23:Rad:M:units}{34.6\mHz}
\setsymbol{LFI:knee:frequency:LFI24:Rad:M:units}{46.2\mHz}
\setsymbol{LFI:knee:frequency:LFI25:Rad:M:units}{24.9\mHz}
\setsymbol{LFI:knee:frequency:LFI26:Rad:M:units}{67.6\mHz}
\setsymbol{LFI:knee:frequency:LFI27:Rad:M:units}{187.4\mHz}
\setsymbol{LFI:knee:frequency:LFI28:Rad:M:units}{122.2\mHz}
\setsymbol{LFI:knee:frequency:LFI18:Rad:S:units}{17.7\mHz}
\setsymbol{LFI:knee:frequency:LFI19:Rad:S:units}{22.0\mHz}
\setsymbol{LFI:knee:frequency:LFI20:Rad:S:units}{8.7\mHz}
\setsymbol{LFI:knee:frequency:LFI21:Rad:S:units}{25.9\mHz}
\setsymbol{LFI:knee:frequency:LFI22:Rad:S:units}{15.8\mHz}
\setsymbol{LFI:knee:frequency:LFI23:Rad:S:units}{129.8\mHz}
\setsymbol{LFI:knee:frequency:LFI24:Rad:S:units}{100.9\mHz}
\setsymbol{LFI:knee:frequency:LFI25:Rad:S:units}{38.9\mHz}
\setsymbol{LFI:knee:frequency:LFI26:Rad:S:units}{58.9\mHz}
\setsymbol{LFI:knee:frequency:LFI27:Rad:S:units}{104.4\mHz}
\setsymbol{LFI:knee:frequency:LFI28:Rad:S:units}{40.7\mHz}

\setsymbol{LFI:knee:frequency:LFI18:Rad:M}{16.3}
\setsymbol{LFI:knee:frequency:LFI19:Rad:M}{15.1}
\setsymbol{LFI:knee:frequency:LFI20:Rad:M}{18.7}
\setsymbol{LFI:knee:frequency:LFI21:Rad:M}{37.2}
\setsymbol{LFI:knee:frequency:LFI22:Rad:M}{12.7}
\setsymbol{LFI:knee:frequency:LFI23:Rad:M}{34.6}
\setsymbol{LFI:knee:frequency:LFI24:Rad:M}{46.2}
\setsymbol{LFI:knee:frequency:LFI25:Rad:M}{24.9}
\setsymbol{LFI:knee:frequency:LFI26:Rad:M}{67.6}
\setsymbol{LFI:knee:frequency:LFI27:Rad:M}{187.4}
\setsymbol{LFI:knee:frequency:LFI28:Rad:M}{122.2}
\setsymbol{LFI:knee:frequency:LFI18:Rad:S}{17.7}
\setsymbol{LFI:knee:frequency:LFI19:Rad:S}{22.0}
\setsymbol{LFI:knee:frequency:LFI20:Rad:S}{8.7}
\setsymbol{LFI:knee:frequency:LFI21:Rad:S}{25.9}
\setsymbol{LFI:knee:frequency:LFI22:Rad:S}{15.8}
\setsymbol{LFI:knee:frequency:LFI23:Rad:S}{129.8}
\setsymbol{LFI:knee:frequency:LFI24:Rad:S}{100.9}
\setsymbol{LFI:knee:frequency:LFI25:Rad:S}{38.9}
\setsymbol{LFI:knee:frequency:LFI26:Rad:S}{58.9}
\setsymbol{LFI:knee:frequency:LFI27:Rad:S}{104.4}
\setsymbol{LFI:knee:frequency:LFI28:Rad:S}{40.7}

\setsymbol{LFI:slope:70GHz:units}{$-1.03$\mHz}
\setsymbol{LFI:slope:44GHz:units}{$-0.89$\mHz}
\setsymbol{LFI:slope:30GHz:units}{$-0.87$\mHz}

\setsymbol{LFI:slope:70GHz}{$-1.03$}
\setsymbol{LFI:slope:44GHz}{$-0.89$}
\setsymbol{LFI:slope:30GHz}{$-0.87$}

\setsymbol{LFI:slope:LFI18:Rad:M:units}{$-1.04$\mHz}
\setsymbol{LFI:slope:LFI19:Rad:M:units}{$-1.09$\mHz}
\setsymbol{LFI:slope:LFI20:Rad:M:units}{$-0.69$\mHz}
\setsymbol{LFI:slope:LFI21:Rad:M:units}{$-1.56$\mHz}
\setsymbol{LFI:slope:LFI22:Rad:M:units}{$-1.01$\mHz}
\setsymbol{LFI:slope:LFI23:Rad:M:units}{$-0.96$\mHz}
\setsymbol{LFI:slope:LFI24:Rad:M:units}{$-0.83$\mHz}
\setsymbol{LFI:slope:LFI25:Rad:M:units}{$-0.91$\mHz}
\setsymbol{LFI:slope:LFI26:Rad:M:units}{$-0.95$\mHz}
\setsymbol{LFI:slope:LFI27:Rad:M:units}{$-0.87$\mHz}
\setsymbol{LFI:slope:LFI28:Rad:M:units}{$-0.88$\mHz}
\setsymbol{LFI:slope:LFI18:Rad:S:units}{$-1.15$\mHz}
\setsymbol{LFI:slope:LFI19:Rad:S:units}{$-1.00$\mHz}
\setsymbol{LFI:slope:LFI20:Rad:S:units}{$-0.95$\mHz}
\setsymbol{LFI:slope:LFI21:Rad:S:units}{$-0.92$\mHz}
\setsymbol{LFI:slope:LFI22:Rad:S:units}{$-1.01$\mHz}
\setsymbol{LFI:slope:LFI23:Rad:S:units}{$-0.95$\mHz}
\setsymbol{LFI:slope:LFI24:Rad:S:units}{$-0.73$\mHz}
\setsymbol{LFI:slope:LFI25:Rad:S:units}{$-1.16$\mHz}
\setsymbol{LFI:slope:LFI26:Rad:S:units}{$-0.79$\mHz}
\setsymbol{LFI:slope:LFI27:Rad:S:units}{$-0.82$\mHz}
\setsymbol{LFI:slope:LFI28:Rad:S:units}{$-0.91$\mHz}

\setsymbol{LFI:slope:LFI18:Rad:M}{$-1.04$}
\setsymbol{LFI:slope:LFI19:Rad:M}{$-1.09$}
\setsymbol{LFI:slope:LFI20:Rad:M}{$-0.69$}
\setsymbol{LFI:slope:LFI21:Rad:M}{$-1.56$}
\setsymbol{LFI:slope:LFI22:Rad:M}{$-1.01$}
\setsymbol{LFI:slope:LFI23:Rad:M}{$-0.96$}
\setsymbol{LFI:slope:LFI24:Rad:M}{$-0.83$}
\setsymbol{LFI:slope:LFI25:Rad:M}{$-0.91$}
\setsymbol{LFI:slope:LFI26:Rad:M}{$-0.95$}
\setsymbol{LFI:slope:LFI27:Rad:M}{$-0.87$}
\setsymbol{LFI:slope:LFI28:Rad:M}{$-0.88$}
\setsymbol{LFI:slope:LFI18:Rad:S}{$-1.15$}
\setsymbol{LFI:slope:LFI19:Rad:S}{$-1.00$}
\setsymbol{LFI:slope:LFI20:Rad:S}{$-0.95$}
\setsymbol{LFI:slope:LFI21:Rad:S}{$-0.92$}
\setsymbol{LFI:slope:LFI22:Rad:S}{$-1.01$}
\setsymbol{LFI:slope:LFI23:Rad:S}{$-0.95$}
\setsymbol{LFI:slope:LFI24:Rad:S}{$-0.73$}
\setsymbol{LFI:slope:LFI25:Rad:S}{$-1.16$}
\setsymbol{LFI:slope:LFI26:Rad:S}{$-0.79$}
\setsymbol{LFI:slope:LFI27:Rad:S}{$-0.82$}
\setsymbol{LFI:slope:LFI28:Rad:S}{$-0.91$}

\setsymbol{LFI:FWHM:70GHz:units}{13\parcm01}
\setsymbol{LFI:FWHM:44GHz:units}{27\parcm92}
\setsymbol{LFI:FWHM:30GHz:units}{32\parcm65}

\setsymbol{LFI:FWHM:70GHz}{13.01}
\setsymbol{LFI:FWHM:44GHz}{27.92}
\setsymbol{LFI:FWHM:30GHz}{32.65}

\setsymbol{LFI:FWHM:LFI18:units}{13\parcm39}
\setsymbol{LFI:FWHM:LFI19:units}{13\parcm01}
\setsymbol{LFI:FWHM:LFI20:units}{12\parcm75}
\setsymbol{LFI:FWHM:LFI21:units}{12\parcm74}
\setsymbol{LFI:FWHM:LFI22:units}{12\parcm87}
\setsymbol{LFI:FWHM:LFI23:units}{13\parcm27}
\setsymbol{LFI:FWHM:LFI24:units}{22\parcm98}
\setsymbol{LFI:FWHM:LFI25:units}{30\parcm46}
\setsymbol{LFI:FWHM:LFI26:units}{30\parcm31}
\setsymbol{LFI:FWHM:LFI27:units}{32\parcm65}
\setsymbol{LFI:FWHM:LFI28:units}{32\parcm66}

\setsymbol{LFI:FWHM:LFI18}{13.39}
\setsymbol{LFI:FWHM:LFI19}{13.01}
\setsymbol{LFI:FWHM:LFI20}{12.75}
\setsymbol{LFI:FWHM:LFI21}{12.74}
\setsymbol{LFI:FWHM:LFI22}{12.87}
\setsymbol{LFI:FWHM:LFI23}{13.27}
\setsymbol{LFI:FWHM:LFI24}{22.98}
\setsymbol{LFI:FWHM:LFI25}{30.46}
\setsymbol{LFI:FWHM:LFI26}{30.31}
\setsymbol{LFI:FWHM:LFI27}{32.65}
\setsymbol{LFI:FWHM:LFI28}{32.66}

\setsymbol{LFI:FWHM:uncertainty:LFI18:units}{0.170\arcm}
\setsymbol{LFI:FWHM:uncertainty:LFI19:units}{0.174\arcm}
\setsymbol{LFI:FWHM:uncertainty:LFI20:units}{0.170\arcm}
\setsymbol{LFI:FWHM:uncertainty:LFI21:units}{0.156\arcm}
\setsymbol{LFI:FWHM:uncertainty:LFI22:units}{0.164\arcm}
\setsymbol{LFI:FWHM:uncertainty:LFI23:units}{0.171\arcm}
\setsymbol{LFI:FWHM:uncertainty:LFI24:units}{0.652\arcm}
\setsymbol{LFI:FWHM:uncertainty:LFI25:units}{1.075\arcm}
\setsymbol{LFI:FWHM:uncertainty:LFI26:units}{1.131\arcm}
\setsymbol{LFI:FWHM:uncertainty:LFI27:units}{1.266\arcm}
\setsymbol{LFI:FWHM:uncertainty:LFI28:units}{1.287\arcm}

\setsymbol{LFI:FWHM:uncertainty:LFI18}{0.170}
\setsymbol{LFI:FWHM:uncertainty:LFI19}{0.174}
\setsymbol{LFI:FWHM:uncertainty:LFI20}{0.170}
\setsymbol{LFI:FWHM:uncertainty:LFI21}{0.156}
\setsymbol{LFI:FWHM:uncertainty:LFI22}{0.164}
\setsymbol{LFI:FWHM:uncertainty:LFI23}{0.171}
\setsymbol{LFI:FWHM:uncertainty:LFI24}{0.652}
\setsymbol{LFI:FWHM:uncertainty:LFI25}{1.075}
\setsymbol{LFI:FWHM:uncertainty:LFI26}{1.131}
\setsymbol{LFI:FWHM:uncertainty:LFI27}{1.266}
\setsymbol{LFI:FWHM:uncertainty:LFI28}{1.287}

\setsymbol{HFI:center:frequency:100GHz:units}{100\,GHz}
\setsymbol{HFI:center:frequency:143GHz:units}{143\,GHz}
\setsymbol{HFI:center:frequency:217GHz:units}{217\,GHz}
\setsymbol{HFI:center:frequency:353GHz:units}{353\,GHz}
\setsymbol{HFI:center:frequency:545GHz:units}{545\,GHz}
\setsymbol{HFI:center:frequency:857GHz:units}{857\,GHz}

\setsymbol{HFI:center:frequency:100GHz}{100}
\setsymbol{HFI:center:frequency:143GHz}{143}
\setsymbol{HFI:center:frequency:217GHz}{217}
\setsymbol{HFI:center:frequency:353GHz}{353}
\setsymbol{HFI:center:frequency:545GHz}{545}
\setsymbol{HFI:center:frequency:857GHz}{857}

\setsymbol{HFI:Ndetectors:100GHz}{8}
\setsymbol{HFI:Ndetectors:143GHz}{11}
\setsymbol{HFI:Ndetectors:217GHz}{12}
\setsymbol{HFI:Ndetectors:353GHz}{12}
\setsymbol{HFI:Ndetectors:545GHz}{3}
\setsymbol{HFI:Ndetectors:857GHz}{4}

\setsymbol{HFI:FWHM:Maps:100GHz:units}{9\parcm88}
\setsymbol{HFI:FWHM:Maps:143GHz:units}{7\parcm18}
\setsymbol{HFI:FWHM:Maps:217GHz:units}{4\parcm87}
\setsymbol{HFI:FWHM:Maps:353GHz:units}{4\parcm65}
\setsymbol{HFI:FWHM:Maps:545GHz:units}{4\parcm72}
\setsymbol{HFI:FWHM:Maps:857GHz:units}{4\parcm39}
\setsymbol{HFI:FWHM:Maps:100GHz}{9.88}
\setsymbol{HFI:FWHM:Maps:143GHz}{7.18}
\setsymbol{HFI:FWHM:Maps:217GHz}{4.87}
\setsymbol{HFI:FWHM:Maps:353GHz}{4.65}
\setsymbol{HFI:FWHM:Maps:545GHz}{4.72}
\setsymbol{HFI:FWHM:Maps:857GHz}{4.39}

\setsymbol{HFI:beam:ellipticity:Maps:100GHz}{1.15}
\setsymbol{HFI:beam:ellipticity:Maps:143GHz}{1.01}
\setsymbol{HFI:beam:ellipticity:Maps:217GHz}{1.06}
\setsymbol{HFI:beam:ellipticity:Maps:353GHz}{1.05}
\setsymbol{HFI:beam:ellipticity:Maps:545GHz}{1.14}
\setsymbol{HFI:beam:ellipticity:Maps:857GHz}{1.19}

\setsymbol{HFI:FWHM:Mars:100GHz:units}{9\parcm37}
\setsymbol{HFI:FWHM:Mars:143GHz:units}{7\parcm04}
\setsymbol{HFI:FWHM:Mars:217GHz:units}{4\parcm68}
\setsymbol{HFI:FWHM:Mars:353GHz:units}{4\parcm43}
\setsymbol{HFI:FWHM:Mars:545GHz:units}{3\parcm80}
\setsymbol{HFI:FWHM:Mars:857GHz:units}{3\parcm67}

\setsymbol{HFI:FWHM:Mars:100GHz}{9.37}
\setsymbol{HFI:FWHM:Mars:143GHz}{7.04}
\setsymbol{HFI:FWHM:Mars:217GHz}{4.68}
\setsymbol{HFI:FWHM:Mars:353GHz}{4.43}
\setsymbol{HFI:FWHM:Mars:545GHz}{3.80}
\setsymbol{HFI:FWHM:Mars:857GHz}{3.67}

\setsymbol{HFI:beam:ellipticity:Mars:100GHz}{1.18}
\setsymbol{HFI:beam:ellipticity:Mars:143GHz}{1.03}
\setsymbol{HFI:beam:ellipticity:Mars:217GHz}{1.14}
\setsymbol{HFI:beam:ellipticity:Mars:353GHz}{1.09}
\setsymbol{HFI:beam:ellipticity:Mars:545GHz}{1.25}
\setsymbol{HFI:beam:ellipticity:Mars:857GHz}{1.03}

\setsymbol{HFI:CMB:relative:calibration:100GHz}{$\lsim 1\%$}
\setsymbol{HFI:CMB:relative:calibration:143GHz}{$\lsim 1\%$}
\setsymbol{HFI:CMB:relative:calibration:217GHz}{$\lsim 1\%$}
\setsymbol{HFI:CMB:relative:calibration:353GHz}{$\lsim 1\%$}
\setsymbol{HFI:CMB:relative:calibration:545GHz}{}
\setsymbol{HFI:CMB:relative:calibration:857GHz}{}

\setsymbol{HFI:CMB:absolute:calibration:100GHz}{$\lsim 2\%$}
\setsymbol{HFI:CMB:absolute:calibration:143GHz}{$\lsim 2\%$}
\setsymbol{HFI:CMB:absolute:calibration:217GHz}{$\lsim 2\%$}
\setsymbol{HFI:CMB:absolute:calibration:353GHz}{$\lsim 2\%$}
\setsymbol{HFI:CMB:absolute:calibration:545GHz}{}
\setsymbol{HFI:CMB:absolute:calibration:857GHz}{}

\setsymbol{HFI:FIRAS:gain:calibration:accuracy:statistical:100GHz}{}
\setsymbol{HFI:FIRAS:gain:calibration:accuracy:statistical:143GHz}{}
\setsymbol{HFI:FIRAS:gain:calibration:accuracy:statistical:217GHz}{}
\setsymbol{HFI:FIRAS:gain:calibration:accuracy:statistical:353GHz}{2.5\%}
\setsymbol{HFI:FIRAS:gain:calibration:accuracy:statistical:545GHz}{1\%}
\setsymbol{HFI:FIRAS:gain:calibration:accuracy:statistical:857GHz}{0.5\%}

\setsymbol{HFI:FIRAS:gain:calibration:accuracy:systematic:100GHz}{}
\setsymbol{HFI:FIRAS:gain:calibration:accuracy:systematic:143GHz}{}
\setsymbol{HFI:FIRAS:gain:calibration:accuracy:systematic:217GHz}{}
\setsymbol{HFI:FIRAS:gain:calibration:accuracy:systematic:353GHz}{}
\setsymbol{HFI:FIRAS:gain:calibration:accuracy:systematic:545GHz}{7\%}
\setsymbol{HFI:FIRAS:gain:calibration:accuracy:systematic:857GHz}{7\%}

\setsymbol{HFI:FIRAS:zero:point:accuracy:100GHz:units}{0.8\MJysr}
\setsymbol{HFI:FIRAS:zero:point:accuracy:143GHz:units}{}
\setsymbol{HFI:FIRAS:zero:point:accuracy:217GHz:units}{}
\setsymbol{HFI:FIRAS:zero:point:accuracy:353GHz:units}{1.4\MJysr}
\setsymbol{HFI:FIRAS:zero:point:accuracy:545GHz:units}{2.2\MJysr}
\setsymbol{HFI:FIRAS:zero:point:accuracy:857GHz:units}{1.7\MJysr}

\setsymbol{HFI:FIRAS:zero:point:accuracy:100GHz}{0.8}
\setsymbol{HFI:FIRAS:zero:point:accuracy:143GHz}{}
\setsymbol{HFI:FIRAS:zero:point:accuracy:217GHz}{}
\setsymbol{HFI:FIRAS:zero:point:accuracy:353GHz}{1.4}
\setsymbol{HFI:FIRAS:zero:point:accuracy:545GHz}{2.2}
\setsymbol{HFI:FIRAS:zero:point:accuracy:857GHz}{1.7}

\setsymbol{HFI:unit:conversion:100GHz:units}{0.2415\MJysrmK}
\setsymbol{HFI:unit:conversion:143GHz:units}{0.3694\MJysrmK}
\setsymbol{HFI:unit:conversion:217GHz:units}{0.4811\MJysrmK}
\setsymbol{HFI:unit:conversion:353GHz:units}{0.2883\MJysrmK}
\setsymbol{HFI:unit:conversion:545GHz:units}{0.05826\MJysrmK}
\setsymbol{HFI:unit:conversion:857GHz:units}{0.002238\MJysrmK}

\setsymbol{HFI:unit:conversion:100GHz}{0.2415}
\setsymbol{HFI:unit:conversion:143GHz}{0.3694}
\setsymbol{HFI:unit:conversion:217GHz}{0.4811}
\setsymbol{HFI:unit:conversion:353GHz}{0.2883}
\setsymbol{HFI:unit:conversion:545GHz}{0.05826}
\setsymbol{HFI:unit:conversion:857GHz}{0.002238}

\setsymbol{HFI:colour:correction:alpha=-2:V1.01:100GHz}{0.9893}
\setsymbol{HFI:colour:correction:alpha=-2:V1.01:143GHz}{0.9759}
\setsymbol{HFI:colour:correction:alpha=-2:V1.01:217GHz}{1.0007}
\setsymbol{HFI:colour:correction:alpha=-2:V1.01:353GHz}{1.0028}
\setsymbol{HFI:colour:correction:alpha=-2:V1.01:545GHz}{1.0019}
\setsymbol{HFI:colour:correction:alpha=-2:V1.01:857GHz}{0.9889}

\setsymbol{HFI:colour:correction:alpha=0:V1.01:100GHz}{1.0008}
\setsymbol{HFI:colour:correction:alpha=0:V1.01:143GHz}{1.0148}
\setsymbol{HFI:colour:correction:alpha=0:V1.01:217GHz}{0.9909}
\setsymbol{HFI:colour:correction:alpha=0:V1.01:353GHz}{0.9888}
\setsymbol{HFI:colour:correction:alpha=0:V1.01:545GHz}{0.9878}
\setsymbol{HFI:colour:correction:alpha=0:V1.01:857GHz}{1.0014}

\providecommand{\sorthelp}[1]{}

\usepackage[nonamebreak]{natbib}
\bibpunct{(}{)}{;}{a}{}{,} %
\usepackage{graphicx}
\usepackage{txfonts}

\usepackage[table,usenames,dvipsnames]{xcolor}

\usepackage[breaklinks, colorlinks, citecolor=blue, colorlinks=true, linkcolor=blue]{hyperref}
\usepackage{fixltx2e} 

\newcommand{\bem}{\begin{multline}}
\newcommand{\eem}{\end{multline}}

\newcommand\be{\begin{equation}}
\newcommand\ee{\end{equation}}
\newcommand{\begm}{\begin{pmatrix}}
\newcommand{\enm}{\end{pmatrix}}

\begin{document}

\title{\Planck\ 2013 results. X. HFI energetic particle effects:
characterization, removal, and simulation}

\author{
\author{\small
Planck Collaboration:
P.~A.~R.~Ade\inst{83}
\and
N.~Aghanim\inst{58}
\and
C.~Armitage-Caplan\inst{87}
\and
M.~Arnaud\inst{70}
\and
M.~Ashdown\inst{67, 6}
\and
F.~Atrio-Barandela\inst{18}
\and
J.~Aumont\inst{58}
\and
C.~Baccigalupi\inst{82}
\and
A.~J.~Banday\inst{90, 10}
\and
R.~B.~Barreiro\inst{64}
% ...
}
\institute{\small
APC, AstroParticule et Cosmologie, Universit\'{e} Paris Diderot, CNRS/IN2P3, CEA/lrfu, Observatoire de Paris, Sorbonne Paris Cit\'{e}, 10, rue Alice Domon et L\'{e}onie Duquet, 75205 Paris Cedex 13, France\\
\and
Aalto University Mets\"{a}hovi Radio Observatory, Mets\"{a}hovintie 114, FIN-02540 Kylm\"{a}l\"{a}, Finland\\
\and
African Institute for Mathematical Sciences, 6-8 Melrose Road, Muizenberg, Cape Town, South Africa\\
\and
Agenzia Spaziale Italiana Science Data Center, Via del Politecnico snc, 00133, Roma, Italy\\
\and
Agenzia Spaziale Italiana, Viale Liegi 26, Roma, Italy\\
\and
Astrophysics Group, Cavendish Laboratory, University of Cambridge, J J Thomson Avenue, Cambridge CB3 0HE, U.K.\\
\and
Astrophysics \& Cosmology Research Unit, School of Mathematics, Statistics \& Computer Science, University of KwaZulu-Natal, Westville Campus, Private Bag X54001, Durban 4000, South Africa\\
\and
Atacama Large Millimeter/submillimeter Array, ALMA Santiago Central Offices, Alonso de Cordova 3107, Vitacura, Casilla 763 0355, Santiago, Chile\\
\and
CITA, University of Toronto, 60 St. George St., Toronto, ON M5S 3H8, Canada\\
\and
CNRS, IRAP, 9 Av. colonel Roche, BP 44346, F-31028 Toulouse cedex 4, France\\
\and
California Institute of Technology, Pasadena, California, U.S.A.\\
\and
Centre for Theoretical Cosmology, DAMTP, University of Cambridge, Wilberforce Road, Cambridge CB3 0WA, U.K.\\
\and
Centro de Estudios de F\'{i}sica del Cosmos de Arag\'{o}n (CEFCA), Plaza San Juan, 1, planta 2, E-44001, Teruel, Spain\\
\and
Computational Cosmology Center, Lawrence Berkeley National Laboratory, Berkeley, California, U.S.A.\\
\and
DSM/Irfu/SPP, CEA-Saclay, F-91191 Gif-sur-Yvette Cedex, France\\
\and
DTU Space, National Space Institute, Technical University of Denmark, Elektrovej 327, DK-2800 Kgs. Lyngby, Denmark\\
\and
D\'{e}partement de Physique Th\'{e}orique, Universit\'{e} de Gen\`{e}ve, 24, Quai E. Ansermet,1211 Gen\`{e}ve 4, Switzerland\\
\and
Departamento de F\'{\i}sica Fundamental, Facultad de Ciencias, Universidad de Salamanca, 37008 Salamanca, Spain\\
\and
Departamento de F\'{\i}sica, Universidad de Oviedo, Avda. Calvo Sotelo s/n, Oviedo, Spain\\
\and
Department of Astronomy and Astrophysics, University of Toronto, 50 Saint George Street, Toronto, Ontario, Canada\\
\and
Department of Astrophysics/IMAPP, Radboud University Nijmegen, P.O. Box 9010, 6500 GL Nijmegen, The Netherlands\\
\and
Department of Electrical Engineering and Computer Sciences, University of California, Berkeley, California, U.S.A.\\
\and
Department of Physics \& Astronomy, University of British Columbia, 6224 Agricultural Road, Vancouver, British Columbia, Canada\\
\and
Department of Physics and Astronomy, Dana and David Dornsife College of Letter, Arts and Sciences, University of Southern California, Los Angeles, CA 90089, U.S.A.\\
\and
Department of Physics and Astronomy, University College London, London WC1E 6BT, U.K.\\
\and
Department of Physics, Florida State University, Keen Physics Building, 77 Chieftan Way, Tallahassee, Florida, U.S.A.\\
\and
Department of Physics, Gustaf H\"{a}llstr\"{o}min katu 2a, University of Helsinki, Helsinki, Finland\\
\and
Department of Physics, Princeton University, Princeton, New Jersey, U.S.A.\\
\and
Department of Physics, University of California, One Shields Avenue, Davis, California, U.S.A.\\
\and
Department of Physics, University of California, Santa Barbara, California, U.S.A.\\
\and
Department of Physics, University of Illinois at Urbana-Champaign, 1110 West Green Street, Urbana, Illinois, U.S.A.\\
\and
Dipartimento di Fisica e Astronomia G. Galilei, Universit\`{a} degli Studi di Padova, via Marzolo 8, 35131 Padova, Italy\\
\and
Dipartimento di Fisica e Scienze della Terra, Universit\`{a} di Ferrara, Via Saragat 1, 44122 Ferrara, Italy\\
\and
Dipartimento di Fisica, Universit\`{a} La Sapienza, P. le A. Moro 2, Roma, Italy\\
\and
Dipartimento di Fisica, Universit\`{a} degli Studi di Milano, Via Celoria, 16, Milano, Italy\\
\and
Dipartimento di Fisica, Universit\`{a} degli Studi di Trieste, via A. Valerio 2, Trieste, Italy\\
\and
Dipartimento di Fisica, Universit\`{a} di Roma Tor Vergata, Via della Ricerca Scientifica, 1, Roma, Italy\\
\and
Discovery Center, Niels Bohr Institute, Blegdamsvej 17, Copenhagen, Denmark\\
\and
European Southern Observatory, ESO Vitacura, Alonso de Cordova 3107, Vitacura, Casilla 19001, Santiago, Chile\\
\and
European Space Agency, ESAC, Planck Science Office, Camino bajo del Castillo, s/n, Urbanizaci\'{o}n Villafranca del Castillo, Villanueva de la Ca\~{n}ada, Madrid, Spain\\
\and
European Space Agency, ESTEC, Keplerlaan 1, 2201 AZ Noordwijk, The Netherlands\\
\and
Helsinki Institute of Physics, Gustaf H\"{a}llstr\"{o}min katu 2, University of Helsinki, Helsinki, Finland\\
\and
INAF - Osservatorio Astrofisico di Catania, Via S. Sofia 78, Catania, Italy\\
\and
INAF - Osservatorio Astronomico di Padova, Vicolo dell'Osservatorio 5, Padova, Italy\\
\and
INAF - Osservatorio Astronomico di Roma, via di Frascati 33, Monte Porzio Catone, Italy\\
\and
INAF - Osservatorio Astronomico di Trieste, Via G.B. Tiepolo 11, Trieste, Italy\\
\and
INAF Istituto di Radioastronomia, Via P. Gobetti 101, 40129 Bologna, Italy\\
\and
INAF/IASF Bologna, Via Gobetti 101, Bologna, Italy\\
\and
INAF/IASF Milano, Via E. Bassini 15, Milano, Italy\\
\and
INFN, Sezione di Bologna, Via Irnerio 46, I-40126, Bologna, Italy\\
\and
INFN, Sezione di Roma 1, Universit\`{a} di Roma Sapienza, Piazzale Aldo Moro 2, 00185, Roma, Italy\\
\and
IPAG: Institut de Plan\'{e}tologie et d'Astrophysique de Grenoble, Universit\'{e} Joseph Fourier, Grenoble 1 / CNRS-INSU, UMR 5274, Grenoble, F-38041, France\\
\and
IUCAA, Post Bag 4, Ganeshkhind, Pune University Campus, Pune 411 007, India\\
\and
Imperial College London, Astrophysics group, Blackett Laboratory, Prince Consort Road, London, SW7 2AZ, U.K.\\
\and
Infrared Processing and Analysis Center, California Institute of Technology, Pasadena, CA 91125, U.S.A.\\
\and
Institut N\'{e}el, CNRS, Universit\'{e} Joseph Fourier Grenoble I, 25 rue des Martyrs, Grenoble, France\\
\and
Institut Universitaire de France, 103, bd Saint-Michel, 75005, Paris, France\\
\and
Institut d'Astrophysique Spatiale, CNRS (UMR8617) Universit\'{e} Paris-Sud 11, B\^{a}timent 121, Orsay, France\\
\and
Institut d'Astrophysique de Paris, CNRS (UMR7095), 98 bis Boulevard Arago, F-75014, Paris, France\\
\and
Institute for Space Sciences, Bucharest-Magurale, Romania\\
\and
Institute of Astronomy and Astrophysics, Academia Sinica, Taipei, Taiwan\\
\and
Institute of Astronomy, University of Cambridge, Madingley Road, Cambridge CB3 0HA, U.K.\\
\and
Institute of Theoretical Astrophysics, University of Oslo, Blindern, Oslo, Norway\\
\and
Instituto de F\'{\i}sica de Cantabria (CSIC-Universidad de Cantabria), Avda. de los Castros s/n, Santander, Spain\\
\and
Jet Propulsion Laboratory, California Institute of Technology, 4800 Oak Grove Drive, Pasadena, California, U.S.A.\\
\and
Jodrell Bank Centre for Astrophysics, Alan Turing Building, School of Physics and Astronomy, The University of Manchester, Oxford Road, Manchester, M13 9PL, U.K.\\
\and
Kavli Institute for Cosmology Cambridge, Madingley Road, Cambridge, CB3 0HA, U.K.\\
\and
LAL, Universit\'{e} Paris-Sud, CNRS/IN2P3, Orsay, France\\
\and
LERMA, CNRS, Observatoire de Paris, 61 Avenue de l'Observatoire, Paris, France\\
\and
Laboratoire AIM, IRFU/Service d'Astrophysique - CEA/DSM - CNRS - Universit\'{e} Paris Diderot, B\^{a}t. 709, CEA-Saclay, F-91191 Gif-sur-Yvette Cedex, France\\
\and
Laboratoire Traitement et Communication de l'Information, CNRS (UMR 5141) and T\'{e}l\'{e}com ParisTech, 46 rue Barrault F-75634 Paris Cedex 13, France\\
\and
Laboratoire de Physique Subatomique et de Cosmologie, Universit\'{e} Joseph Fourier Grenoble I, CNRS/IN2P3, Institut National Polytechnique de Grenoble, 53 rue des Martyrs, 38026 Grenoble cedex, France\\
\and
Laboratoire de Physique Th\'{e}orique, Universit\'{e} Paris-Sud 11 \& CNRS, B\^{a}timent 210, 91405 Orsay, France\\
\and
Lawrence Berkeley National Laboratory, Berkeley, California, U.S.A.\\
\and
Max-Planck-Institut f\"{u}r Astrophysik, Karl-Schwarzschild-Str. 1, 85741 Garching, Germany\\
\and
McGill Physics, Ernest Rutherford Physics Building, McGill University, 3600 rue University, Montr\'{e}al, QC, H3A 2T8, Canada\\
\and
National University of Ireland, Department of Experimental Physics, Maynooth, Co. Kildare, Ireland\\
\and
Niels Bohr Institute, Blegdamsvej 17, Copenhagen, Denmark\\
\and
Observational Cosmology, Mail Stop 367-17, California Institute of Technology, Pasadena, CA, 91125, U.S.A.\\
\and
Optical Science Laboratory, University College London, Gower Street, London, U.K.\\
\and
SB-ITP-LPPC, EPFL, CH-1015, Lausanne, Switzerland\\
\and
SISSA, Astrophysics Sector, via Bonomea 265, 34136, Trieste, Italy\\
\and
School of Physics and Astronomy, Cardiff University, Queens Buildings, The Parade, Cardiff, CF24 3AA, U.K.\\
\and
Space Sciences Laboratory, University of California, Berkeley, California, U.S.A.\\
\and
Special Astrophysical Observatory, Russian Academy of Sciences, Nizhnij Arkhyz, Zelenchukskiy region, Karachai-Cherkessian Republic, 369167, Russia\\
\and
Stanford University, Dept of Physics, Varian Physics Bldg, 382 Via Pueblo Mall, Stanford, California, U.S.A.\\
\and
Sub-Department of Astrophysics, University of Oxford, Keble Road, Oxford OX1 3RH, U.K.\\
\and
Theory Division, PH-TH, CERN, CH-1211, Geneva 23, Switzerland\\
\and
UPMC Univ Paris 06, UMR7095, 98 bis Boulevard Arago, F-75014, Paris, France\\
\and
Universit\'{e} de Toulouse, UPS-OMP, IRAP, F-31028 Toulouse cedex 4, France\\
% ...
}

}
%

 \date{Received XX, 2013; accepted XX, 2014}



 \abstract{ We describe the detection, interpretation, and removal of
   the signal resulting from interactions of high energy particles
   with the \Planck\ High Frequency Instrument (HFI). There
   are two types of interactions: heating of the 0.1\,K
   bolometer plate; and glitches in each detector time stream.
   The transient responses to detector glitch shapes are not simple single-pole 
   exponential decays and fall into three families. The glitch
   shape for each family has been characterized empirically in flight
   data and these shapes have been used to remove glitches from the
   detector time streams.  The spectrum of the count rate per unit
   energy is computed for each family and a correspondence is made to
   the location on the detector of the particle hit.  Most of the
   detected glitches are from Galactic protons incident on the die
   frame supporting the micro-machined bolometric detectors. In the \Planck\ orbit at L2,
   the particle flux is around $5\,{\rm cm}^{-2}\,{\rm s}^{-1}$ and is
   dominated by protons incident on the spacecraft with energy
   $>$39\,MeV, at a rate of typically one event per second per
   detector. Different categories of glitches have different
   signatures in the time stream. Two of the glitch types have a low
   amplitude component that decays over nearly 1\,s. This
   component produces excess noise if not properly removed from the
   time-ordered data.  We have used a glitch detection and subtraction
   method based on the joint fit of population templates. The
   application of this novel glitch subtraction method removes excess
   noise from the time streams. Using realistic simulations, we find that
   this method does not introduce signal bias into the \Planck\
   data. }

 \keywords{cosmology: observations -- cosmic background radiation --
   surveys -- space vehicles: instruments -- instrumentation:
   detectors -- cosmic rays}

 \authorrunning{Planck Collaboration} \titlerunning{HFI energetic
   particle particle effects: characterization, removal and
   simulations}
   \maketitle

\section{Introduction}

\all2013resultspapers

This paper, one of a set associated with the 2013 release of data from
the \Planck\footnote{\Planck\ (\url{http://www.esa.int/Planck}) is a
  project of the European Space Agency (ESA) with instruments provided
  by two scientific consortia funded by ESA member states (in
  particular the lead countries France and Italy), with contributions
  from NASA (USA) and telescope reflectors provided by a collaboration
  between ESA and a scientific consortium led and funded by Denmark.}
mission \citep{planck2013-p01}, describes the detection of high energy
particle impacts on the 0.1\,K stage and bolometric detectors in the
\Planck\ High Frequency Instrument \citep[HFI,][]{Lamarre2010} and
removal of these systematic effects from the millimetre and
submillimetre signals.  A full description of the HFI can be found
elsewhere
\citep{planck2011-1.5,planck2011-1.7,planck2013-p03,planck2013-p03c,planck2013-p03f,planck2013-p03d}.

\begin{figure}
  \centering
  \includegraphics[width=1\columnwidth]{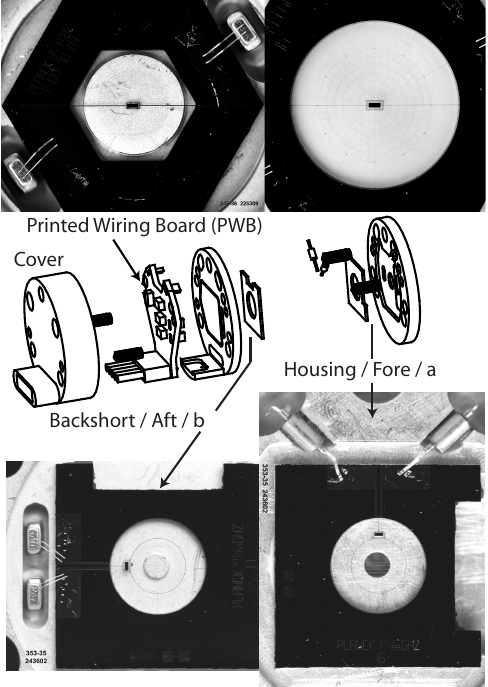}
  \caption{\label{bolometer_figure}\textit{Top left and right}: Completed
    multimode \citep[see][]{maffei2010}
    545\,GHz and single-mode 143\,GHz SWB bolometer modules.
    \textit{Middle}: An exploded view of the assembly of a PSB (showing the
    definition of the ``a'' and ``b'' detectors of the pair).
    Alignment pins,
    shown in solid black, fix the aft and fore bolometer assemblies to
    an angular precision of $<0.1^\circ$.  The SWB assembly is similar to
    the PSB aft bolometer assembly and does not include a feedhorn
    aperture integrated with the module package.  \textit{Bottom}: PSB pair
    epoxied in the module parts prior to mating.  To the right, the
    feedhorn aperture can be seen through the fore bolometer in the
    housing. To the left, the quarter-wave backshort can be seen through
    the aft bolometer absorber mesh.}
\end{figure}

Bolometers \citep{holmes2008}, such as those used in HFI on \Planck,
are phonon-mediated thermal detectors with finite response time to
changes in the absorbed optical power.  The bolometer consists of a
micro-machined silicon nitride (Si-N) mesh absorber with a germanium
(Ge) thermistor suspended from the Si die frame. Each bolometer is
mounted on a metal housing, and these are assembled into two types of
bolometer modules, as shown in Fig.~\ref{bolometer_figure}: a spider web
bolometer \citep[SWB,][]{Bock1995} which detects total power; and a
polarization sensitive bolometer \citep[PSB,][]{jones07}. In the PSB module
there are two bolometers to independently measure power in each of the
two linear polarizations. The
bolometers are mounted on a copper-plated stainless steel plate cooled
to 0.1\,K and stabilized within a few microkelvin of the temperature set
point. Two ``dark'' bolometer modules are blanked off at the 0.1\,K
plate and are used to monitor systematic effects.  The 0.1\,K
bolometer plate is surrounded by a roughly 1.5\,mm thick aluminium box
cooled to 4.5\,K.  Light is coupled into each bolometer module using a
feedhorn at 4.5\,K and a filter stack at 1.5\,K \citep{Lamarre2010}.

At the Earth-Sun Lagrange point L2, high energy particles from the Sun
and Galactic sources -- primarily protons, electrons and helium nuclei --
are incident on the spacecraft. This particle flux causes two main
effects in HFI that have been reported previously. There is a
time-variable thermal load on the 0.1\,K plate \citep{planck2011-1.3}
and a significant rate of glitches in the bolometric signal
\citep{planck2011-1.7}. In this paper, we report on the evolution of
these effects over the entire period of HFI operations, from 3 July
2009 to 14 January 2012, and describe the analysis technique used to
remove the effect of glitches from the data.  Three families of
glitches, ``long,'' ``short,'' and ``slow'' \citep[or ``longer'' as
named in][]{planck2011-1.7}, were found by comparing and stacking many
events.  Their characteristics are detailed in
Sect.~\ref{sec:families}. Long glitches dominate the overall counts
and exhibit an additional slow decay with a time constant of the order
of 2\,s, requiring template subtraction from the data. The new
analysis, presented in Sect.~\ref{sec:method}, employs a
joint fitting of the three glitch family templates and removal of the
slow tails. In Sects.~\ref{methodres} and \ref{simustudy} we show
that our adopted technique improves the noise performance of HFI. We
have simulated the effect of the glitches on the data quality and find
that systematic biases are small, contributing $<10^{-4}$ to the
cosmological power spectrum.

We also present the coincident counts, energy distribution, total
counts, and variations of glitch shapes within each family. These data,
taken together with the beam-line and laboratory tests using spare HFI
bolometers \citep{particletest, planck2011-1.3}, allow identification
of the physical cause of the glitch events for two of the types (Sect.~\ref{sec:characinterp}).
Long glitches are due to energy
absorption events in the Si die and short glitches are due to events in the
optical absorbing grid or Ge thermistor. The cause of the slow
glitches, however, has not been identified. We also describe some rare effects not previously
reported, including the response of the instrument to solar flares,
secondary showers, and large high-energy events.

\section{Glitch families}
\label{sec:families}

For instruments with bolometric detectors fielded on balloons
\citep{Crill03, Masi06, archeops1, BLASTprocess} and space missions
\citep[see][for \textit{Herschel}-SPIRE]{SPIRE}, the glitches are flagged in the data time stream and
then the flagged data are omitted. For most of these instruments, the
transfer function is deconvolved from the time stream prior to
flagging glitches. This has the benefit of increasing the signal-to-noise ratio
 on each glitch, which minimizes the fraction of falsely flagged
data.

\begin{figure}
  \centerline{\includegraphics[width=\columnwidth]{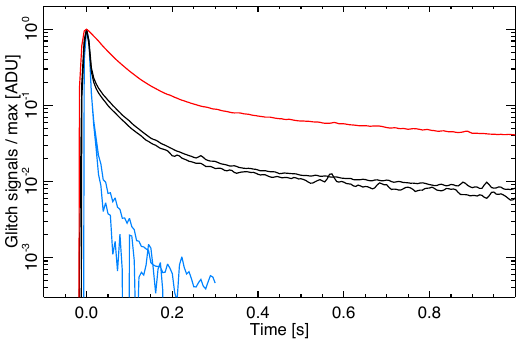}}
  \caption{Examples of three distinct families of glitch transfer functions
    for a typical PSB-a bolometer. Events like 
    the blue curves are called ``short'' events, those like the
    black curves are called ``long,'' and those like the red curves are called
    ``slow.'' Typical variation of the shape within each family is
    shown for short and long glitches. The differences between long
    glitch shapes are modelled by a single nonlinearity parameter
    relating the amplitude of the slow tail of events with their peak
    amplitude. There is no apparent glitch tail associated with short
    events.}
\label{typeGL}
\end{figure}

\begin{figure}
  \centerline{\includegraphics[width=\columnwidth]{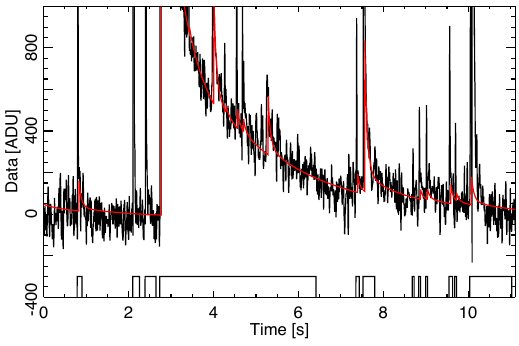}}
  \caption{\textit{Black}: Segment of raw data for one detector at 143\,GHz before any
    deglitching; an estimate of the sky signal has been
    subtracted. \textit{Red}: A time stream reconstructed from the
    estimated templates of long glitches with the method presented in
    Sect.~\ref{sec:method}. We have chosen a region in the vicinity of
    a large event. Data that are flagged for the analysis are
    indicated by the lines at the bottom of the figure. Notice the
    high level of confusion between long glitch signals.}
\label{fig:Timeline}
\end{figure}

The population of glitches in \Planck\ HFI is unusual. The glitch rate
of $\sim 1$\,s$^{-1}$ per detector is significantly higher than in the other
instruments. Also, three distinct families of glitch transfer
functions, shown in Fig.~\ref{typeGL}, are present in the data. The
transfer function of each glitch family is described by a sum of three
or four decaying exponential terms. For a given bolometer, the families,
called short, long, and slow, are differentiated by the relative
amplitudes of each term. Only the short glitches have a transfer
function matching the instrument transfer function. The long glitches
dominate the total rate and have a significant amplitude above the
noise level, with a time constant $>1$\,s as illustrated in
Fig.~\ref{typeGL}. Deconvolving the instrument transfer function prior
to glitch flagging did not improve glitch detection, and the large
amplitude at times $>1$\,s, which is the typical time between glitches,
means there is glitch pile up. The high rate of long glitches leads to
considerable confusion of glitch signals, as can be seen in
Fig.~\ref{fig:Timeline}, and makes it difficult to clean glitches in
\Planck\ HFI data.

Omitting flagged data would lead to omission of $>90\%$ of
the data, so  we have developed a strategy to fit and to remove
precise templates of long glitches from the time-ordered
data. We take advantage of the excellent pointing accuracy
\citep{planck2011-1.1} and redundancy of the scan strategy, and have
adopted a method that iterates between estimates of the signal from
the sky and signal from the glitches \citep{planck2011-1.7}. A key to
the effectiveness of this method, presented in Sect.~\ref{sec:method},
is a comprehensive understanding of the three families of glitches and a
determination of the stability of these families over the course of
the mission.  In this section, we describe the features of
the different glitch  families  that are relevant for their removal. Identification
and classification of the glitches has been done using previous
versions of the method, as introduced in \cite{planck2013-p03c}. This
was an iterative process, since accurate glitch property determination
is necessary to ``tune'' the method for effective detection and
cleaning, and an efficient method is necessary to separate glitches
into families and derive properties.  Templates of each of the three
glitch families are obtained from stacking many normalized glitches,
and the general features are derived from fitting the normalized decay
profile with a sum of exponentials. The methodology is detailed in
Sect.~\ref{templates}.

\subsection{Short glitches}

\begin{figure}
\centering
\includegraphics[width=1.\columnwidth]{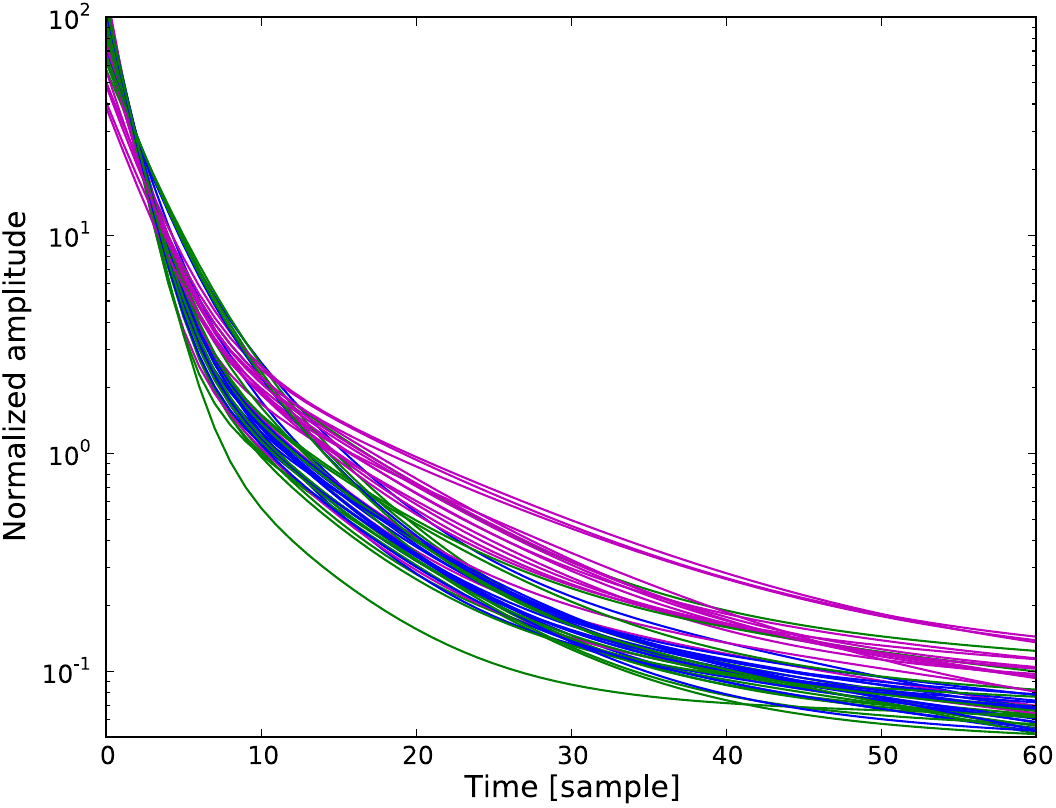}
\caption{Short glitch templates for all detectors obtained using the
  method discussed in Sect.~\ref{templates}. Blue lines are for PSB-a,
  green for PSB-b, and magenta for SWB (see Fig.~\ref{typeGL} for the
  definition of the ``a'' and ``b'' bolometers in a PSB pair). One
  sample corresponds to 5\,ms.}
\label{fig:short_templates}
\end{figure}

The highest-amplitude events are the short
glitches.  The short glitch time response, or template, is shown in
Fig.~\ref{fig:short_templates}. Short events have a rise time 
much shorter than the sampling period of 5\,ms, followed by a fast
decay. These rapid variations can cause oscillations in the
electronics, and the response amplitude depends on the precise moment
of the glitch within the sampling period. So throughout the paper, the
amplitude of the glitch is defined as the peak value of the sampled
data. The shortest exponential decay component has a time constant
similar to the fast part of the optical transfer function; three
additional terms (corresponding to the tail) have very low
amplitudes ($<10^{-2}$ of the peak), and time constants of
typically 40\,ms, 400\,ms, and 2\,s.  The 2\,s decaying
signal has an amplitude about $5 \times
10^{-5}$ of the peak and was detected only after stacking short glitches measured
over the entire mission. The transfer function of short glitches
nearly matches the instrument optical transfer function.

\begin{figure}
  \centering
  \includegraphics[width=\columnwidth]{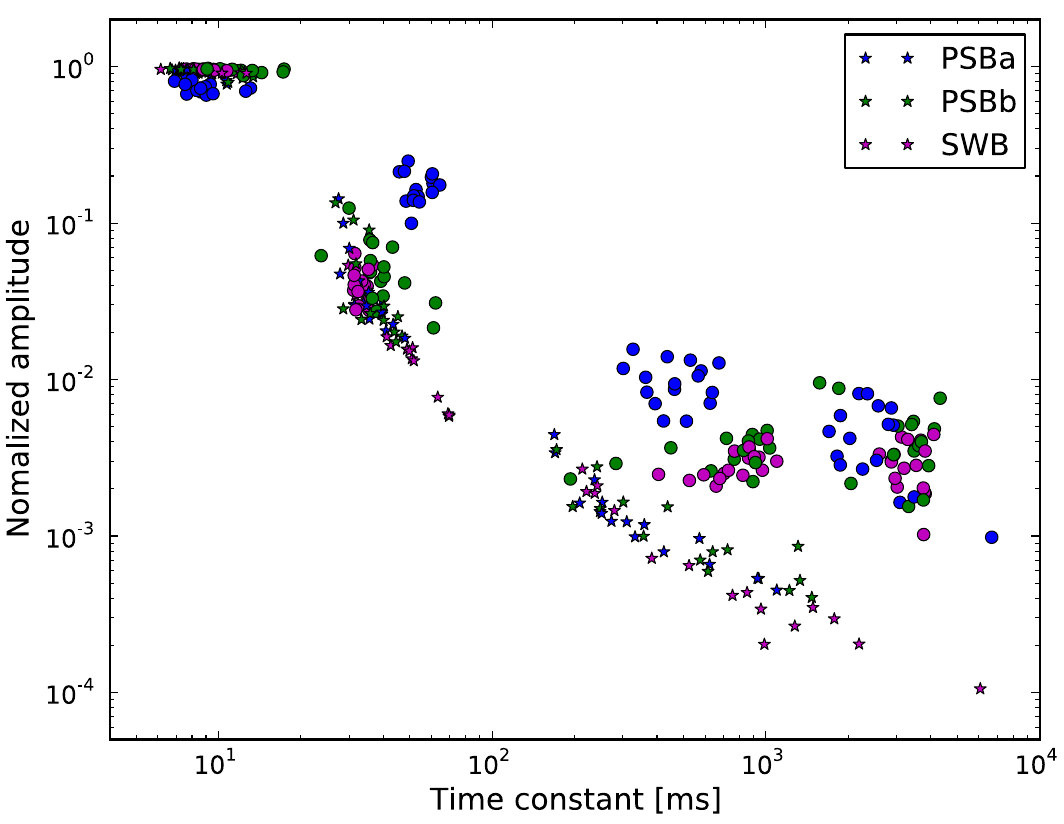}
  \caption{\label{Template_numbers} Parameters of the glitch templates
    built from the sum of four decaying exponentials for long glitches
    and three exponentials for short glitches. Fitted amplitudes
    versus time constants for all exponentials are displayed for all
    bolometers. Stars indicate short glitches and circles indicate long
    glitches. The type of bolometer is indicated by colour (blue is
    PSB-a, green is PSB-b, and magenta is SWB). Values plotted are
    obtained after fitting exponentials on templates estimated by
    stacking a large number of events and normalized to one at the
    peak. A three-point filter was applied to the data prior to the
    fit of the exponentials (see text).}
\end{figure}

The time constants and amplitudes derived by fitting
three exponentials to short glitch measurements are given in
Fig.~\ref{Template_numbers} for all the bolometers. The fit has been
performed after stacking events with amplitudes between 1200 and 2400
times the noise (around 2\,keV of deposited energy), which corresponds
to energies of the transition between the two subcategories of short
events. The 2\,s time constant is not included in the template fit.
We have verified that,
given the counts of short glitches, and assuming a random distribution
of glitches, the 2\,s tails produce a signal that is three orders
of magnitude lower than the white noise, and so can be neglected in
the data processing. We observe some scatter of the values across
bolometers, although this may reflect the degeneracies in the fit
of the different time constants and amplitudes rather than variations of the
template shape between bolometers. A careful study of the degeneracies
in the parameter fitting is presented in the analysis of the transfer
function \citep{planck2013-p03c}.

The amplitude of the tail is variable from event to event, but is
small enough that its variability does not affect the processing of
data. Moreover, we have found the existence of two subcategories of
short glitches (one event of each subcategory is shown in
Fig.~\ref{typeGL}), which can be distinguished by the amplitude of the
tails. In fact, the tail has smaller amplitude for higher energy
events and larger amplitude for lower energy events, with a variation
of a factor of about two.

\subsection{Long glitches}
\label{sub:longgl}

\begin{figure}
\centering
\includegraphics[width=1.\columnwidth]{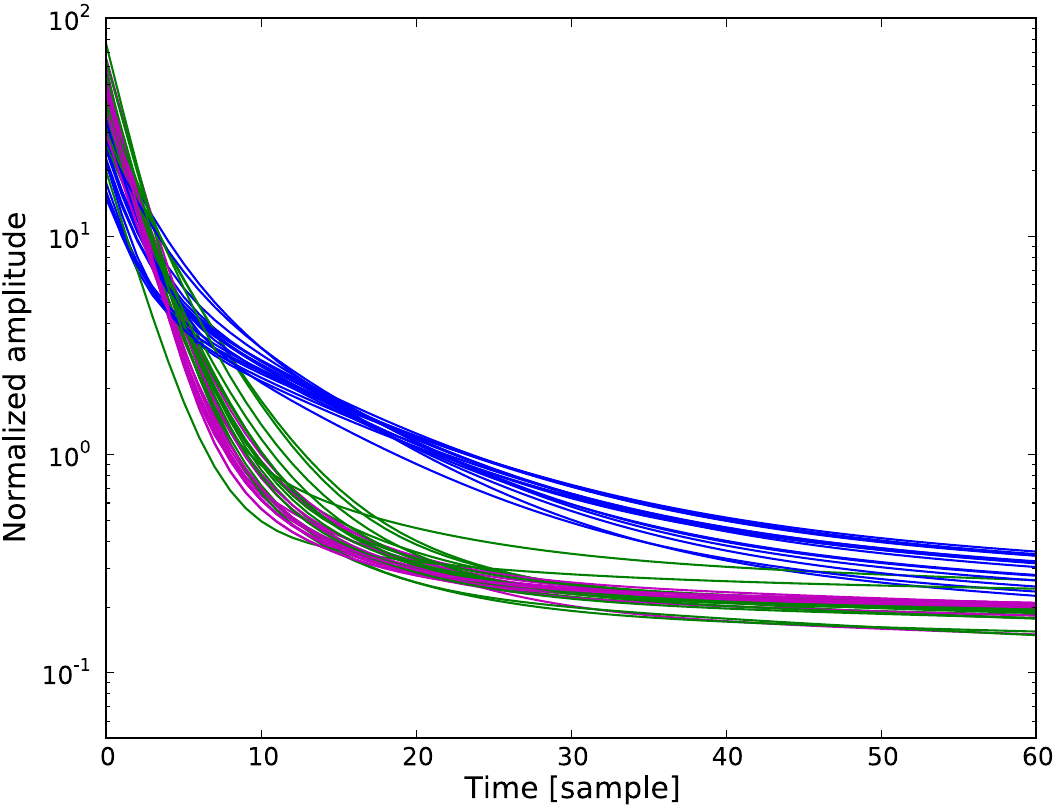}
\caption{Long glitch templates for all detectors.  Blue is for
  PSB-a, green for PSB-b and magenta for SWB.}
\label{fig:long_templates}
\end{figure}

The long glitches are, on average,  of lower amplitude than short
glitches, but their rate is more than an order of magnitude  greater.
 Most of the glitches in the HFI data are these long events. Their
intermediate and slow decays are responsible for  a noise
excess at frequencies between 0.001 and 1\,Hz (see
Fig.~\ref{fig:spres}), and they must be subtracted in order to
reach the expected noise level.
The rise time is very
fast, much less than the sampling period, like the short glitches.
The shortest exponential decay has a time constant similar to the fast
part of the optical transfer function, followed by a tail with a much
larger amplitude than we see for the short glitches.  In
Fig.~\ref{fig:long_templates} long glitch templates are shown for
several bolometers. As with the short glitches, they are estimated by
stacking a large number of events, and fitted with a sum of four
exponential terms.

The slow tail has typical time constants of 50\,ms, 500\,ms, and 2\,s
for PSB-a bolometers and 35\,ms, 500\,ms and 2\,s for PSB-b and SWB
bolometers. The component with intermediate time constant (35 to 50\,ms) starts with
an amplitude relative to the peak of around 6\% for PSB-a and around
3\% for PSB-b and SWB bolometers.  The components with long time constants 
(500\,ms and 2\,s) have an amplitude that is about 0.2\% of the event
peak.

The most striking feature in Fig.~\ref{fig:long_templates} is that the
component with intermediate time constant decays faster and has lower 
amplitude for PSB-b and SWB than for PSB-a.  The
component with the longest time constant, $\sim2$\,s,
 has a similar amplitude for all bolometers. The time constants
and amplitudes are shown in Fig.~\ref{Template_numbers} for all the
bolometers.

We do not observe significant changes in the time constants of the
long glitch tail from event to event.  However, the amplitude
of the long tail relative to the fast part does show significant
dispersion (see Sect.~\ref{sec:interplg}).
The quoted amplitudes of the long and intermediate tails
are for lower energy events, but we find that the
amplitude of the tail of long glitches (relative to the peak) is a
function of the peak amplitude of the events. This nonlinearity can be
described empirically to good accuracy with a simple linear
function that is fitted and accounted for in the processing. The
nonlinearity corresponds to a factor of two at most between the lowest
and highest energies of detected events. This factor is lower for
PSB-b and SWB bolometers. This is illustrated by Fig.~\ref{typeGL},
which shows two typical long events at different amplitudes, both
normalized to the peak. The difference in amplitude of
the slow tail is clearly visible.

Finally, in a small proportion of events the ratio of amplitude of the fast
part to that of the slow part  is different from the majority of long
glitches. In particular, slightly less than 10\% of events have a
smaller amplitude tail than long glitches and about 0.5\% have a
higher tail (not including slow glitches). Those events do not fit
into any of the categories of glitches, but their slow decaying tails, after
$\approx 20$\,ms from the peak amplitude, are similar to the long
glitches. Because of this, they are implicitly accounted for in
the glitch removal procedure.

\subsection{Slow glitches}
\label{slowgl}

\begin{figure}
  \centerline{\includegraphics[width=\columnwidth]{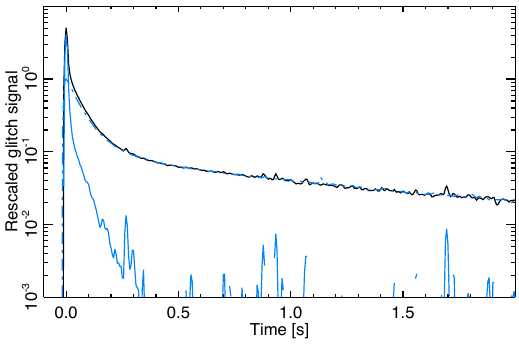}}
  \caption{Comparison of a long glitch (black) and a slow glitch (dot-dashed
    blue). For this plot, the two high energy events were rescaled to
    match after 200\,ms. The difference is shown in blue.}
\label{fig:compsnlg}
\end{figure}

Slow glitches are detected only for PSB-a. They have a rise time
comparable to the optical time constant, i.e., much slower than the
two other glitch families. The tail of a slow glitch is similar to
that of a long glitch.  Fig.~\ref{fig:compsnlg} shows a comparison
of a high amplitude long glitch event with a slow glitch. The
intermediate time constant of long glitches, which is of the order of
50\,ms for PSB-a, is slightly (but significantly) larger for slow
glitches, and this time constant is the fastest for those events, as
can be seen in Fig.~\ref{fig:compsnlg}. The two tails are proportional
to good accuracy from about 200\,ms after the peak
amplitude. Even with these differences, the long glitch template for a
given bolometer is a good proxy for the slow glitches in the same
bolometer.

\subsection{Population counts}
\label{popcounts}

\begin{figure}
  \centerline{\includegraphics[width=1.0\columnwidth]{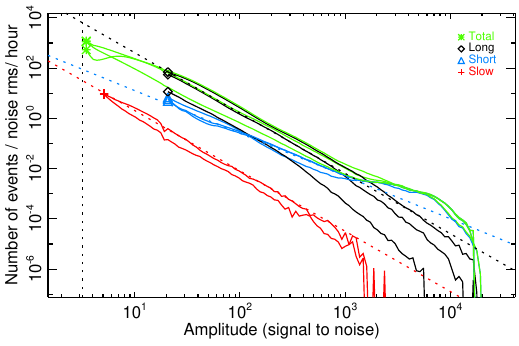}}
  \caption{Distributions, $dN/dE$, of the three families of glitches
    with respect to the peak amplitude, in signal-to-noise ratio units, for
    three different bolometers: 143-1a; 143-2a; and 217-1. The blue line is for
    the short glitch population, black is for long, red is for slow,
    and green is for total. We have chosen
    bolometers with very different behaviour: 143-1a has one of the lowest
    glitch rates while 143-2a and 217-1 have high glitch rates. 
    The faint-end break of the glitch counts is visible for
    217-1, a bolometer that has no slow glitches (see
    Sect.~\ref{slowgl}). Power laws are shown for comparison as dashed
    lines. Indices and amplitudes of power laws are chosen to match
    the distribution of  bolometer 143-2a. Indices are
    $-2.4$ for long and slow glitches, and $-1.7$ for short glitches. The
    vertical dashed line indicates the detection threshold. There is no
    attempt here to separate the long and short glitch populations below 20
    times the rms noise. The slow glitch spectrum is shown down to 5
    times the rms.}
\label{fig:PopSp}
\end{figure}

Figure~\ref{fig:PopSp} shows the distribution $dN/dE$ of the three
populations of glitches for one detector as a function of the amplitude of events in
signal-to-noise ratio units. The type of each event
is determined by the method of
Sect.~\ref{sec:method}.
We can see that long glitch events are dominant at lower amplitudes,
while short events dominate at higher amplitudes.  The separation
between short and long events is not efficient below about
$20\,\sigma$ (i.e., 20 times the rms noise), because the tails of long
glitches, used to distinguish between the two types of events, are
barely detectable. Nevertheless, we will see later that long glitches
are dominant at lower amplitudes, as can be guessed by extrapolating
the counts. Long glitches are dominant in the overall counts. The
distribution of long glitches is well fitted by a steep power law of
index $-2.4$ between typical amplitudes of 20 and 1000 times the rms
noise (with some variations among
bolometers), and shows breaks at both the bright and faint ends. The
faint-end break is close to the glitch detection threshold, which
is fixed at $3.2\,\sigma$, and is detected very significantly (as we
will discuss in Sect.~\ref{sec:interplg}). The
submillimetre channels (545 and 857\,GHz) are more sensitive to long
glitches. For those detectors, we observe a peak in the differential
counts at energies close to the detection threshold. We
will see in Sect.~\ref{sec:interplg} that very few events are expected
below the detection threshold. We observe a large scatter from
bolometer to bolometer in the amplitude scaling of the long glitch distribution,
but the overall shape is preserved.

The distribution of short glitches has a bump at amplitudes around
$3000\,\sigma$ for all bolometers from 100 to 353\,GHz. There
is a clear break at very high amplitudes, which is mostly due to the
nonlinearity of the detectors at those amplitudes and to the
saturation of the analogue-to-digital convertor \citep[ADC, see ][for
a discussion of the effect on the data]{planck2013-p03}.
The bump in the distribution is
not apparent in the submillimetre detectors, but this is due to the
nonlinearity smearing it out.  For amplitudes below about
$1000\,\sigma$, the distribution of short glitches is well represented
by a power law with a typical index of $-1.7$. Short events with an
energy corresponding to the bump appear to have shorter timescales
than events corresponding to the power law.

The shapes of the distributions of slow and long glitches are very
similar. We see in Fig.~\ref{fig:PopSp} that long glitches are more
numerous than slow glitches, but the distributions shown are
normalized to the peaks of events. After rescaling on the amplitude of
the slow tails, which are similar for the two populations, the slow
glitch distributions end up much closer to the long glitch
distributions. The slow glitches are then contributing very
significantly to the noise power excess.

\begin{figure}
  \centering
  \includegraphics[width=1\columnwidth]{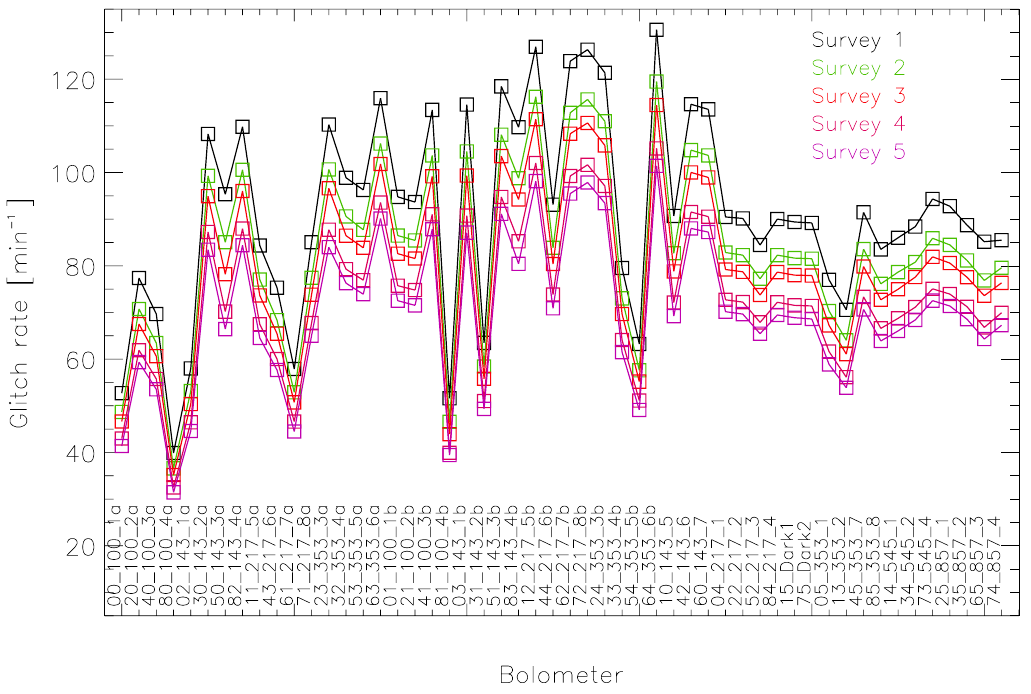}
  \caption{\label{rate_survey}Glitch rates for each
    bolometer. Points represent the mean values during each
    6 month survey.}
\end{figure}

The glitch rates per bolometer are given in Fig.~\ref{rate_survey} for
each six-month survey.  Numbers are derived by integrating
differential counts down to the detection threshold. Long glitches
dominate the counts. Variations across bolometers are mainly explained
by the differences in the sensitivity to long glitches. In particular,
channels with lower rates also have more events below the detection
threshold.
 As described in~\cite{holmes2008}, bolometers in PSBs
were selected from die on the same wafer. Also, attempts were made to
select die from neighboring locations on that same wafer. In 14 of 16
PSBs, the PSB-a and -b are from the same wafer. Due to limited yield,
for 2 of the 100GHz PSBs, 100-2 and 100-4, PSB-a and PSB-b are from
different wafers. Nevertheless, as shown in Fig.~\ref{rate_survey}
there is no significant difference between the bolometers in these two
PSBs compared to difference in bolometers in PSBs where both
bolometers are from the same wafer.

\section{Glitch detection and removal}

Now that we have described the three glitch families, we
focus on the detection and removal method.  An earlier 
version was described  in
\cite{planck2011-1.7}. Here, we summarize our approach and note the
changes that have been implemented. We also demonstrate the efficiency
of the method and show the impact on the noise power spectra.

\subsection{Method}
\label{sec:method}

The technique for glitch removal
iterates between sky signal estimation at the ring level
and glitch detection and fitting.  The iteration procedure, and the method used to estimate
the sky signal for the first iteration, are presented in \citet{planck2011-1.7}.
Each ring is analysed independently \citep[see][for the definition of a
ring]{planck2011-1.1}. A simple digital three-point filter (0.25, 0.5,
0.25) is used for glitch detection. This filter performs similarly to 
an optimal matched filter based on fast glitch
shapes, since the bolometer time constant is very close to the
sampling period. It has the additional advantage of reducing ringing
around large events resulting from the electronic filter
\citep{Lamarre2010} and demodulation of the data. Depending on the
ring, five to nine iterations between sky signal estimation and
deglitching are necessary for convergence. 

\subsubsection{Glitch detection and template subtraction}

After estimating the sky signal (as described in Sect.~\ref{skysubtr})
at the previous iteration, and removing it from the data, events are
detected by selecting local maxima in the three-point filtered data
above a noise threshold set at $3.2\,\sigma$ on a sliding window of
1000 samples.\footnote{The choice of $3.2\,\sigma$ is a compromise
  between false event detection and completeness of glitch detection
  \citep[][see figure~7]{planck2011-1.7}.} On each window, a joint fit
of the amplitudes of short and long glitch templates is performed
simultaneously for all detected events in the window, centred at the
maximum event. The short and long glitch templates are built
independently, and each is a sum of exponentials, three terms for
short and four terms for long glitches.  Templates are estimated by
stacking events detected and separeted into categories by earlier
versions of the same method. The parameters of the templates are
essentially independent of the changes in the method, since templates
are determined using the list of detected high energy glitches, which
is not sensitive to the details of the processing, and it was not
necessary to iterate the glitch template estimation.  The method used
to estimate the templates is described in Sect.~\ref{templates}.  The
joint template fitting of both short and long glitches is a
significant improvement over the method used previously
\citep{planck2011-1.7}. It has improved the accuracy of determination
of the long glitch template amplitude and led to smaller glitch
residuals in the final noise. Only one overall amplitude is used as a
free parameter for each glitch type and for each event.

Detected glitches are fit to the templates starting from three samples
after the maximum for small-amplitude glitches in PSB-b or SWB
detectors, increasing to eight samples after the maximum for large-amplitude 
glitches, and from six to eight samples in PSB-a detectors
(because of the presence of slow glitches).  In order to break
degeneracies between template amplitudes in the case of high
confusion between events, we use the prior that amplitudes are
positive. The parameter fitting is performed by analytical
minimization of $\chi^2$, as detailed in Eqn.~2 of
\cite{planck2011-1.7}. Two parameters per event are fitted: 
the amplitudes of the long and short glitch tails described by the two
templates. Priors are implemented by setting parameters that fall outside the allowed range to fixed values; in particular, 
negative amplitudes are replaced by zeros, and then another
minimization is performed.

Slow glitches are detected as long glitches with extremely large
amplitudes. Thus such events are processed in the same
way as long glitches but with higher fitted amplitudes relative to
the peak. When a slow event is detected, another iteration of the
$\chi^2$ minimization is performed, in which the long template is
adapted by removing the fastest exponential.

If the measured long glitch template amplitude for an event in the
vicinity of the largest event is above the threshold (fixed to 0.5
times the expected amplitude for a long glitch at the measured peak
amplitude), then the long glitch template is removed from the
data. The expected amplitude of the long glitch tail accounts for
the nonlinearity described in Sect.~\ref{sub:longgl}; we use
a simple empirical quadratic law, which is fitted to the data,
relating the glitch template amplitude and the peak amplitude.  To
remove the template from the data, we use the fitted value of the
amplitude of the long glitch template for events above
$10\,\sigma$. Glitches with overall amplitude below $10\,\sigma$ are
treated either as long or slow glitches, and the amplitude is fixed to
the expected values.  This is because the long glitches dominate at
low amplitude, and because the fitting errors are larger than the model
uncertainties. The fitted short glitch template is not removed from
the data (this has no impact on the data as discussed later): affected data are
just flagged.

\subsubsection{Data flagging}

For long and slow events, data are flagged in the vicinity of the peak when the fitted long glitch template is above three
times the rms noise, and the flagged samples are not used for
scientific analysis (or for fitting the next event).  For events
belonging to the short category, or at least detected as such, i.e., for
which the fitted long glitch template amplitude is below the detection
threshold, data for which the short glitch template is above
$0.1\,\sigma$ are flagged.

At the end of the glitch detection and subtraction procedure, a
matched filter, optimized to detect 2\,s exponential decays, is
applied to the data, in order to find events that are missed by the
method, or events for which the fit failed. This introduces an extra
flagging of about 0.1\% of the data. Additional glitch cleaning and
the use of different adaptive filters were not found not to improve
the data any further.

\subsubsection{Sky subtraction}
\label{skysubtr}

The sky signal is estimated at each iteration on each fixed pointing
period (ring), detector by detector, and this is done independently
using the redundancies of the measurements.  Third-order spline
coefficients are fit to nodes separated by 1\farcm5 to account for
subpixel variations, which could otherwise be detected as glitch
signals. Flags set during previous iterations are used to reject
data, and estimates of the long glitch templates are removed prior to the sky
signal estimation.

Special sky estimation and glitch removal procedures are required when
scanning through or near bright objects, including planets and bright
sources in the Galactic plane.  This is to account for systematic
errors in the subtraction of the sky signal that could be falsely
flagged as glitches.  These errors are mainly due to two effects: slow
cross-scan variations of the pointing within the ring scan time,
yielding pointing uncertainties of the order of a few seconds of arc
\citep{planck2013-p03}; and subpixel variations that are
not entirely captured by the spline coefficients.

For rings with large signal in the Galactic plane, we add to the noise
variance the square of a term representing the signal reconstruction error,
which is 1\% of the signal for
submillimetre channels (545 and 857\,GHz) and 0.5\% for the other channels. This
automatically increases the threshold used for glitch detection for
bright regions of the sky.  Also, long glitch templates are not
removed for high sky signals. This correction of the threshold is
effective for all rings for the submillimetre channels,
but only applies to a few percent of the data close to the Galactic
centre for the other channels \citep[see][]{planck2011-1.7}. Further special treatment is
required to process the data on bright
planets. The motions of planets during one hour on the sky
(which are mainly cross-scan) and the cross-scan pointing
uncertainties are both accounted for. We fit the amplitude of the
planet signal for each scan, rescale the averaged signal estimate
on the ring using the estimated amplitude coefficient to reconstruct
the sky signal, and remove the best estimate of it from the 
data. This procedure neglects main-lobe beam asymmetries. However, we
find that it is a good approximation at this stage, given the small
cross-scan shift of the planet signal during the time taken to scan
one ring.

Sky signal estimation and glitch detection and subtraction on each ring are
iterated until the results converge to sufficient accuracy. For most rings, six iterations are
necessary \citep[see][]{planck2011-1.7}.

\subsubsection{Template estimation}
\label{templates}

\begin{figure*}[!ttt]
  \centerline{\includegraphics[width=2.0\columnwidth]{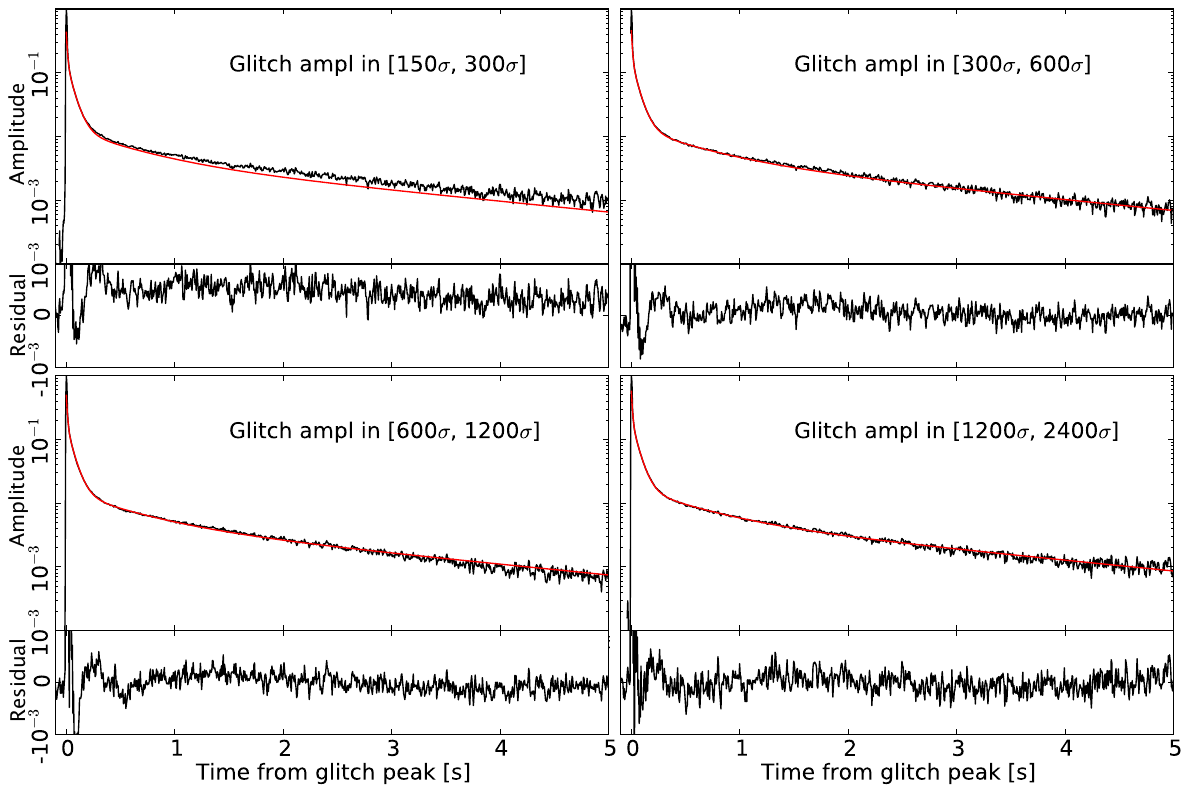}}
  \caption{Stacking of long glitches of bolometer 143-3a in different
    energy windows, after normalizing each event to the maximum
    amplitude. The solid lines show the long glitch template obtained
    after fitting four exponentials, using the data shown in the lower
    right panel. Templates are rescaled in amplitude in each pannel to
    match the stacked data due to the nonlinearity of the long glitch
    tail relative to the peak presented in Sect.~\ref{sub:longgl}. In
    each panel the lower curve shows the difference between the
    stacked data and the template. There is good agreement between the
    template and the stacked data, typically within 1\%.}
  \label{fig:stacktmp}
\end{figure*}

Long and short glitch templates, built from a sum of exponentials, are
fitted to stacked (normalized and averaged) large amplitude events
(typically 1000 times the rms noise). Any representation of the glitch
pulses providing a good fit to the stacked glitch data would have been
convenient for the analysis. However, the functional form is motivated
by very simple physical thermal models of the bolometer and its
environment. Time series are not used directly because of noise
contribution to the long tail which is filtered by the model.  The
stacked data are produced using the median value for each bin
corresponding to the sampling period index after the maximum. The
median is used to avoid significant contribution from other glitches.
Figure~\ref{fig:stacktmp} shows the difference between long glitch
stacks and the best-fit template (with four exponentials) in different
energy windows for one bolometer (143-3a).  The glitch template is
determined using events with amplitudes between 1200 and
$2400\,\sigma$.

The difference between the stacked glitch data and the fitted template
is small, but not consistent with zero, reflecting the limit of the
summed exponential model. The difference is systematic, has a low
frequency component, and is about 1\% of the template, which is much
smaller than the fitting errors in each individual glitch described in
Sect.~\ref{GLcontamin} (see Fig.~\ref{fig:resnoisering} in
particular).  We therefore expect that the impact of this inaccuracy
on the final results is negligible.  We see no evidence for
significant variations of the shape of the template with the amplitude
of the glitch, over a wide range of amplitudes: the difference between
the glitch template and the stacks computed over different ranges of
glitch amplitude, between 300 and $5000\,\sigma$ (after correcting for
the nonlinearity of the glitch tail amplitude relative to the peak
described in Sect.~\ref{sub:longgl}), is no larger than in the
amplitude range of 1200--$2400\,\sigma$ where the template is fitted.
The template used for the analysis has been estimated for a limited
range of energies to avoid mismatch due to the nonlinearity effects.

For small glitches, with amplitudes $<300\,\sigma$, there is a
non-negligible difference between the stacked glitches and the fitted
template.  This is attributed to strong selection biases, e.g.,
Eddington-type bias, which affects the stacked signal, and not to real
variations of the glitch template. This bias is clearly observed for
stacked samples, showing a difference in the mean offset before the
events compared with after the events (measured a few tenths of a
second later, when the glitch tail is negligible). This bias is
expected to be small for the high-amplitude events  used for 
modelling the templates.

\subsection{Results}
\label{methodres}

\begin{figure}
  \centerline{\includegraphics[width=\columnwidth]{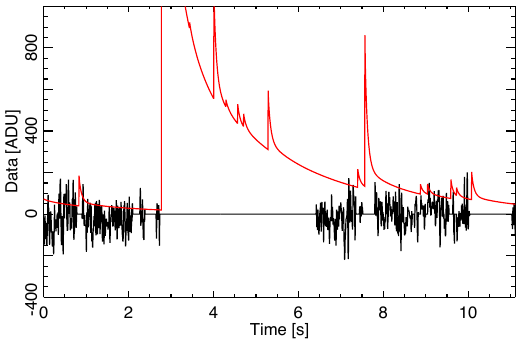}}
  \caption{Segment of data for one detector at 143\,GHz after long
    glitch template subtraction. The same segment before template
    subtraction and flagging is shown in Fig.~\ref{fig:Timeline}. An
    estimate of the sky signal has been subtracted. The time stream reconstructed from the estimated templates of
    long glitches that have been subtracted from the data is shown in
    red. The flagged data that are not used for scientific
    analysis are set to zero.}
\label{fig:ResTimeline}
\end{figure}

The fraction of the data that are found to be contaminated by glitches and are ignored in the analysis varies
from 6\% (for detectors with fewer long glitches) to 20\%, depending
on the bolometer. The averaged fraction of flagged data by HFI
frequency is 14.4, 16.1, 16.8, 17.1, 12.8, and 11.2 \%, for 100 to 857
GHz.  Figure~\ref{fig:ResTimeline} shows a segment of
data for a bolometer with a high glitch rate after glitch
template subtraction. This is the same segment as shown in
Fig.~\ref{fig:Timeline} before deglitching. It shows clearly
that the template subtraction is very effective.

\begin{figure*}
  \centering
  \includegraphics[width=1.0\columnwidth]{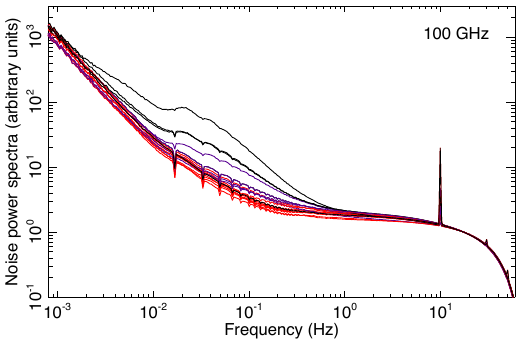}
  \includegraphics[width=1.0\columnwidth]{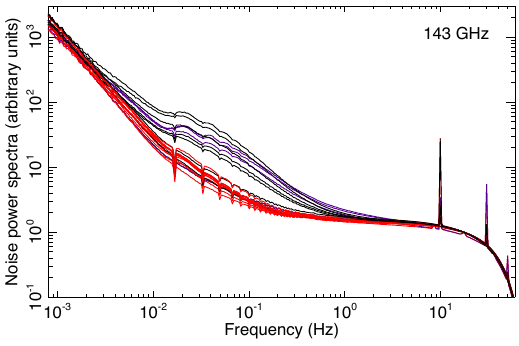}
  \includegraphics[width=1.0\columnwidth]{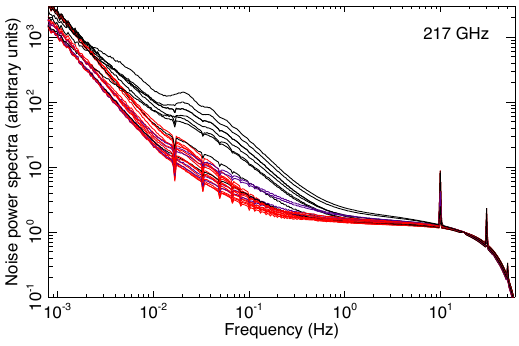}
  \includegraphics[width=1.0\columnwidth]{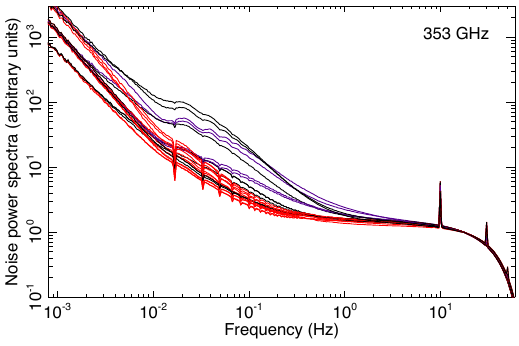}
  \caption{Power spectra of the component of noise that does not
    project onto the sky, for each bolometer at frequencies between
    100\,GHz (top left panel) and 353\,GHz (bottom right panel).
    Black curves correspond to the power spectra without subtraction
    of templates for PSB-b and SWB detectors, while the blue curves
    are for PSB-a detectors. Power spectra are rescaled in amplitude
    so that they all match at 20\,Hz. The red curves are with template
    subtraction.  Power spectra are computed over subsets of about 100
    rings and are averaged over the nominal mission. The sky signal
    has been previously subtracted from the data. This creates the narrow dips in the spectra at the harmonics
    of the spin frequency. Data have also been three-point
    filtered and flagged after glitch detection.  The large spikes in
    the spectra, unsubtracted at this stage, are induced by the $4\,$K
    J-T cooler and are described in \cite{planck2013-p03}. The
    roll-off of the spectra above 10\,Hz is mainly due to the
    three-point filter. Glitches contribute to the
    power spectra between 0.002\,Hz and 2\,Hz. The power spectra of
    residual contamination is significantly reduced after glitch
    template subtraction, at frequencies below a few hertz. There is
    much less dispersion between the power spectra from different
    detectors after correction -- less
    than a factor of two -- indicating that contamination from
    remaining glitches is not a dominant effect.}
  \label{fig:spres}
\end{figure*}

The subtraction of long and slow glitch templates significantly
reduces the contamination of the cosmological and astrophysical
signal.  Figure~\ref{fig:spres} compares the power spectra of noise
with and without glitch template subtraction. The power spectra are
computed after sky subtraction (only the part of the noise that does
not project onto the sky is left), from raw three-point filtered data
(same filter as applied to the data prior to sky signal estimation and
deglitching), for all the detectors from 100 to 353\,GHz that are used
for cosmology. Both data sets were flagged in the same way after
glitch detection, gaps were filled with white noise plus low frequency
noise estimated from smoothing data with a Gaussian. Power spectra are
computed on data chunks of size corresponding to about 100 rings,
which are averaged over the period covered by the nominal mission.  An
estimate of the sky signal has been previously removed ring by ring
for each bolometer using redundancies, as described in
Sect.~\ref{sec:method}, and after subtracting the averaged signal for
each ring, to avoid filtering the noise on timescales larger than the
ring scale.
Before correction, the power spectra are highly contaminated by glitches
at frequencies between 0.002\,Hz and 2\,Hz. The 2\,s tail is
responsible for the excess between 0.01 and 0.1\,Hz, while the
intermediate time constant of 50\,ms produces an excess around
0.4\,Hz, explaining why PSB-a detectors have a higher relative excess
at that frequency. We observe a large reduction of the  residual
noise after template subtraction.  Some detectors have very low
contamination from long glitches; this is the case for 143-5 and
143-1a, which correspond to the two lowest spectra before correction
(top right panel of Fig.~\ref{fig:spres}).  The
improvement after deglitching is effective but small for these
detectors, as expected, and their corrected power spectra are very
close to the fundamental limit of the noise. The power spectra for
these detectors are then indicative of the power spectra of noise
without glitches.The power spectra for the other detectors approach
 this limit after template subtraction. Noting
that the $1/f$ part of the spectrum varies in amplitude from bolometer
to bolometer, we conclude that residual contamination from
glitches is below the noise level at all scales for all detectors.

One and only one of the 100\,GHz bolometers (100-1a) has an extra
family of glitches that are not properly accounted for by our
deglitching method.  Some of these events are detected as ``long'' in
the processing, and then the long glitch template is incorrectly
subtracted (those events are shorter than the long events). But the
matched filter, designed to detect long glitch tails (either positive
or negative) allows us to flag segments of data when this happens,
limiting the effect. This bolometer is the one
with the largest excess noise power at frequencies around 0.1\,Hz. This is still a
small effect, which is accounted for in the total noise budget
\citep{planck2013-p03}.

The power spectra of noise after sky subtraction are
indicative of the quality of the glitch correction, but cannot
be used to derive accurately the contribution to the maps. The
reason is that part of the noise is filtered due to the ``noisy''
estimation of templates by the fitting procedure, which are then
removed from the data. This is not the case for the signal projecting
onto the sky, since the method iterates between sky signal and glitch
detection, and then template fitting and subtraction. The contribution
of glitches to the total power after data reduction at the map level,
and at the ring level, can be studied with simulations (Sect.~\ref{simustudy}).

\begin{figure}
  \centering
  \includegraphics[width=1.0\columnwidth]{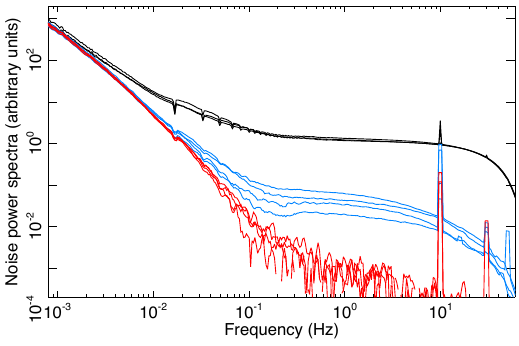}
  \caption{Cross- and auto-power spectra between detectors at
    100\,GHz. Spectra are computed in the same way as described in the
    caption of Fig.~\ref{fig:spres}. Black curves correspond to some
    of the auto-spectra shown in the same figure.  Data have been
    rescaled so that auto-spectra match at 20\,Hz, and so that the
    white noise level is approximately unity. Blue curves correspond
    to cross-spectra between detectors from the same pair. Red curves
    are for detectors from different pairs. The extra correlation
    observed for PSB pair bolometers, above $0.1$\,Hz and at the level
    of 2 to 7\% of the white noise spectrum, is due to residual
    glitches below the threshold. Some residual correlation at the
    level of about 0.1\% at 0.2\,Hz can be seen between other
    bolometer pairs that are spatially close to each other; this might
    result from imperfect sky signal subtraction due to pointing
    drift, since the error is coherent between different nearby
    detectors. The low frequency noise is highly correlated between
    detectors and is due to thermal fluctuations of the focal
    plane. The level of correlation of the 4K lines in the plot is not
    representative of the residual obtained after the TOI processing.}
  \label{crossspect}
\end{figure}

As we will see in Sect.~\ref{sec:interplg}, long glitches occur
simultaneously in both bolometers of a PSB pair. Consequently,
glitches below the detection threshold, or any unsubtracted tails, are
expected to induce residual additional correlations between bolometer
time streams from the same pair. In order to evaluate the contribution
from residual glitches we have computed the cross-power spectra
between bolometers. These are shown in Fig.~\ref{crossspect} for
detectors at 100\,GHz, after averaging cross- and auto-power spectra
computed on data segments of about 100 rings over the entire nominal
mission. The sky signal has been removed and data have been processed
in the same way as previously described for auto-spectra.
Bolometers from the same PSB pair have an extra correlation above
0.2\,Hz, which is around 2--7\% of the white noise level in the
spectrum for 100\,GHz bolometers. The correlation is 2--3\% for
PSB pairs at 143\,GHz, 1--4\% at 217\,GHz, and 2--4\% at
353\,GHz. Simulations presented in the next section show that the
residual noise excess is of the order of 5\%
(Fig.~\ref{fig:resnoisering}); given our modelling of glitches
(Sect.~\ref{sec:interplg}), and noting that about
half of the glitches have the same amplitude in PSB-a and PSB-b
detectors (see Sect.~\ref{sec:interplg}), we should have roughly 5\%
correlation in the noise spectrum. This is slightly higher than
the 3\% observed for this pair. The difference could be due to
uncertainties in our modelling and model extrapolations. It is
important to note that we observe stronger correlations in pairs for which both
detectors have few detected long glitches, e.g., the 100-4a/4b pair,
for which the detectors have the lowest glitch rates (as shown in
Fig.~\ref{rate_survey}), and which have 7\% cross-correlation (as seen
in Fig.~\ref{crossspect}), the highest observed for any pair. This is
entirely explained by the fact that both detectors then have more
glitches below the detection threshold, since the counts for long
glitches are similar between detectors, but are essentially scaled in
energy, as described in Sect.~\ref{sec:interplg}.  By contrast,
 pairs for which the faint-end break in counts is above
the detection threshold have significantly lower correlations above
0.1\,Hz.

We have shown that glitches below the detection threshold are
responsible for the extra correlation in the noise by averaging the
two time streams from each pair of bolometers, and looking for
events above $3.2\,\sigma$ (of the noise) in the combined
data. This allows us to detect coincident glitches with amplitudes
lower by a factor of $\sqrt 2$ for about half of the events, enabling
us to divide the correlated noise by a factor of 2, with only
0.1\% of additional data flagging.  This shows, without ambiguity,
that the extra correlation between two bolometers in PSB pairs is due
to undetected long glitches below the threshold.

Furthermore, this analysis confirms that the number of
undetected glitches is small and that a change in slope (or break) must
 happen in the counts close to the detection threshold, as we would
otherwise have expected more correlations in the noise between PSB
pairs. Those glitches then account for a small fraction of the total
noise power.

\section{Impact of glitch residuals on final results}
\label{simustudy}
\subsection{Simulations}

In order to estimate the contamination from glitches that remains after
processing, we have performed simulations of glitch time streams,
incorporating the glitch properties that we measure in
data. The simulations include the following features.
\begin{itemize}
\item Glitches are generated using a Poisson distribution, with
  sub-sample resolution.
\item Generated glitches follow the population spectra that are found
  in data for each population by combining bolometers, and using the
  model explained in Sect.~\ref{sec:characinterp}. Population spectra
  are rescaled in energy for each bolometer to match the measured
  counts. In particular, we use to build the model the measured number
  counts of long glitches in the submillimetre channels because those
  channel are more sensitive to low-energy glitches, below the
  faint-end break. Population spectra of short and slow glitches are
  extrapolated at low energy (for which events can not be detected
  individually) using power laws.
\item The nonlinearity of the slow tail of long glitches relative to
  the peak amplitude is included (see Sect.~\ref{sub:longgl} for
  details).
\item The temporal shape of glitches is simulated using templates for
  the slow parts (after about 20\,ms), analytically correcting the
  effect of the sampling average and of the three-point filter on the
  amplitude of the exponentials, since those are applied to the data
  before template estimation. The fast part of glitches is simulated
  using subsample resolution, and then averaged over the sample
  period with a simple boxcar average.
\end{itemize}
The simulations do not include the apparent scatter of the slow tail
amplitude of long glitches, apart from the intrinsic scatter due to
the variations of the arrival time in the sample period.
Those fluctuations do not strongly
affect the performance of the method, since the slow part of the
glitches is fitted before subtraction, and so the amplitude
variations are absorbed by the fit. Moreover, we used a fixed template
for short glitches, which is not completely realistic, since we have
observed scatter in the slow tail, and we have identified two
subcategories of short glitches. The effect of using a single, fixed
template for short glitches should be very small, since the short
glitch templates are not removed. We did not include the 2\,s
exponential decay for short glitches in the simulations, since the
uncertainty in its amplitude is large. Nevertheless, we estimate
that the contribution to the power spectrum should
be below 0.1\% of the noise power spectrum, given the distribution of
short glitches. The effects of nonlinearity of the bolometers and the
ADC are not simulated. The nonlinearity has the effect of
reducing the fast part of the glitches, which  dominates at very
high amplitudes. We do not expect this to degrade the performance of
the method for the slow part of the glitches, since the template fitting
starts a few samples after the peak, in a part of the signal that is
not affected by nonlinearity. Finally, we do not include the
correlation of glitches between PSB pairs, but treat each bolometer
independently.

\begin{figure}
  \centerline{\includegraphics[width=1.0\columnwidth]{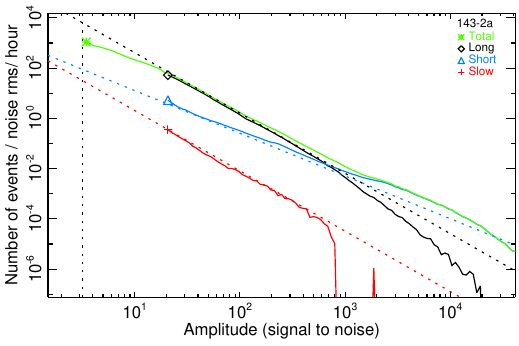}}
  \caption{Differential counts for the three populations of glitches
    for one of the detectors with a high rate of glitches (143-2a)
    measured from simulated data. This can be compared to the results on real data
    shown in Fig.~\ref{fig:PopSp} for the same bolometer. Dashed lines
    are shown for reference and are identical to the lines shown for
    the real data. There is almost perfect agreement between
    simulations and data, except for very high amplitude glitches
    $\gsim 10^4 \sigma$. Those glitches are affected by the bolometer
    nonlinearity in the real data, an effect that we did not simulate
    at this stage, explaining the discrepancies in the counts. We
    simulated slightly fewer very high amplitude slow glitches than in
    the \Planck\ data. This affects only about one event per day.}
  \label{fig:inspectra}
\end{figure}

To the generated glitch signal, we add Gaussian noise containing a
white component and a low frequency component described by a power law
fitted to the data (with an index and $f_{\rm knee}$ parameter). We
also add a time stream of pure signal obtained by scanning over a
simulated cosmic microwave background (CMB)
map, as well as Galactic dust and point source maps from
the \Planck\ Sky Model \citep{delabrouille2013}, using the pointing
solution derived for the actual data. The constructed signal
time stream is formed by interpolating the extracted signal from the
map to limit subpixel effects \citep[see][for the
methodology]{reinecke2006}. We also filter the simulated data using
the same three-point filter as used for the data. The differential
counts recovered from the simulations after processing are shown in
Fig.~\ref{fig:inspectra} for bolometer 143-2a.
There is very good agreement with the spectra recovered in \Planck\
bolometer data, shown in Fig.~\ref{fig:PopSp}.

\subsection{Error estimation}
\label{errorestim}

\subsubsection{Evaluation of signal bias due to the deglitching procedure}

Due to the high signal-to-noise ratio of HFI data, the sky signal
subtraction is a critical part of the deglitching procedure. Errors in
the sky signal estimation could easily induce spurious detection of
glitches and errors in the template subtraction, which would then
correlate with the signal. This could bias the sky signal estimation
for two main reasons: first by flagging data as bad slightly more
often on average when the sky signal is higher (or lower); and second,
by subtracting slightly more glitch template signals when the
fluctuations of the sky signal are of a given sign.

We have verified the absence of bias in the signal with the help of
simulations.  Specifically, we have computed signal rings by
projecting pure signal time streams used for simulations,
which we write as ${r_0}_i(p)$, where $p$ is the ring pixel and
$i$ is the ring number.  We have also projected the
simulated observed data from which we have subtracted estimated glitch
templates, after applying the deglitching procedure used for
real data and rejecting flagged data. We write this last quantity as ${\hat r}_i(p)$ for data
ring $i$. We then computed the average binned power spectrum of
${r_0}_i(p)$ as:
\begin{equation}
  P_0(q) = {1\over N N_q} \sum_{i=1}^N
  \sum_{k\in D_q}{r_0}_i(k)\,{r_0}_i(k)^\dagger,
\end{equation}
where $\dagger$ denotes the transpose conjugate, ${r_0}_i(k)$ is the
Fourier transform of ${r_0}_i(p)$, $D_q$ and $N_q$ are the intervals in
$k$ and the number of modes in bin $q$, respectively, and we have
summed over $N$ rings (with $N$ of the order of 10\,000). We also
computed the average cross-power spectrum between ${r_0}_i(p)$ and
${\hat r}_i(p)$ as
\begin{equation}
P_{\rm c}(q) = {1\over N N_q} \sum_{i=1}^N
 \sum_{k\in D_q}{r_0}_i(k)\,{\hat r}_i(k)^\dagger.
\end{equation}
In the absence of bias in the signal introduced by any of the two
effects previously described, $P_{\rm c}(q)$ should be an unbiased
estimate of $P_0(q)$, i.e.,
\begin{equation}
P_0(q) = \left\langle P_{\rm c}(q)\right\rangle,
\end{equation}
where $\left\langle \cdot \right\rangle$ is the ensemble average over
an infinite number of realizations of noise and glitches.

\begin{figure}[!ttt]
  \centerline{\includegraphics[width=1.0\columnwidth]{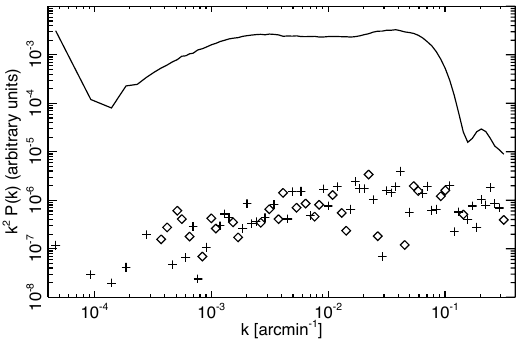}}
  \caption{Constraints on the signal bias after deglitching. The black
    curve is the power spectrum ($k^2 P_0(k)$,
    with $k$ taken at the centre of each bin $q$) of ring data after
    projecting a pure simulated signal time stream for one 
    detectors, containing CMB anisotropies,
    Galactic dust and point source signal, and using the actual
    pointing.  For the other displayed points we computed the
    cross-spectrum between: (1) the estimated signal on rings after
    subtracting long glitch templates from the simulated data and
    after flagging; and (2) the input signal only averaged on the same
    rings. Crosses and diamonds correspond to the difference between
    this cross-spectrum and the pure input signal power spectrum shown
    in black, for positive and negative points, respectively. The
    quantity displayed is $k^2 (P_{\rm c}(k)-P_0(k))$. We averaged the power spectra of 10\,000
    rings. Simulations were performed for detector 143-2a which has a high
    glitch rate.
    In the absence of bias in
    the signal due to the deglitching, we expect that the cross-power
    spectrum is an unbiased estimator of the input
    signal power spectra, and that the difference should then be
    compatible with zero. We do not detect significant bias in any of
    the 100 logarithmically spaced bins, and we place an upper limit
    of $5\times 10^{-4}$ on each bin at the scales relevant for CMB
    analysis.
  }
  \label{fig:evalbias}
\end{figure}

In practice, the quantities $P_0(q)$ and $P_{\rm c}(q) - P_0(q)$ are
evaluated for simulations of the 143-2a detector for $N = 10\,000$
rings. We selected rings with no strong Galactic signal, to avoid
giving too much weight to the Galaxy, keeping about $95\%$ of the
rings. Results are shown in Fig.~\ref{fig:evalbias} for
logarithmically spaced bins.
We can see that $P_{\rm c}(q) - P_0(q)$ is compatible with zero at all
scales. We estimate an upper limit to the bias of around $5\times
10^{-4}$ in individual bins at all scales relevant for cosmology,
i.e., for $k$ corresponding to $1\leq\ell\leq2000$. The absence of
significant bias due to the deglitching procedure is verified at the
ring level, but this conclusion can be drawn at the map level, and
hence at the CMB power spectrum level. Also, we do not expect to
observe differences for the CMB measured with other bolometers, since
the presence of bias has to do with the errors made in the sky signal
reconstruction, and not on errors made in glitch template fitting and
subtraction.  Nevertheless, we have performed the same exercise for an
SWB bolometer at 143\,GHz (143-5). We did not find any bias and can
place an even lower limit of $2\times 10^{-4}$, due to the lower rate
of glitches for this bolometer.

The situation is different for strong Galactic signals and for high
frequency channels, since the detection threshold is increased with
the signal amplitude and even long glitch templates are not
subtracted for very high sky signal, as described in
Sect.~\ref{sec:method}. We expect this to slightly bias the estimation
of the sky signal in the positive direction, since less positive
glitch signal is removed in regions of strong sky signal. This effect
is described in detail in \cite{planck2011-1.7}. In particular we
observed an effect of the order of $4 \times 10^{-4}$ of the signal
amplitude at 545\,GHz. The effect on the beam response estimation is
studied in detail in \cite{planck2013-p03c}.

Because the sky signal is estimated using data from individual rings,
the estimate is noisy, with an rms that is reduced by
roughly a factor of 7 compared to the rms of the noise in each
measurement. This induces some weak correlations in the glitch
detection between each circle in the ring, after the sky signal has
been removed from the data.  However, this effect does not bias the
estimation of the signal, as  shown above, but correlates
slightly, at the level of 1\%, with the remaining noise (after
flagging and template subtraction) between subsets of data measured by
taking half rings.  We have studied this effect in
\cite[][see Table~2 and Fig.~32]{planck2013-p03} and it has been taken into account for the noise
prediction.

\subsubsection{Residual glitch contamination}
\label{GLcontamin}

\begin{figure*}
  \centering
  \includegraphics[width=1.0\columnwidth]{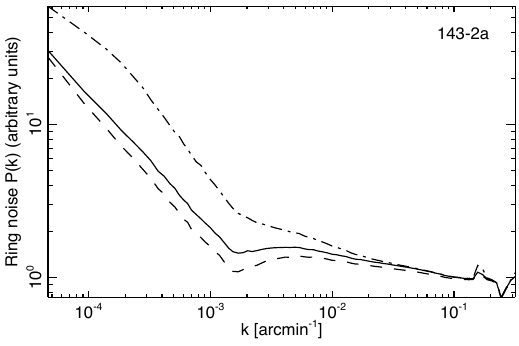}
  \includegraphics[width=1.0\columnwidth]{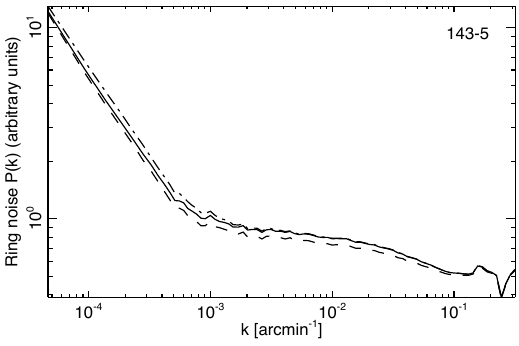}
  \caption{Estimation of the power spectra of noise residuals after
    averaging data on rings for the 143-2a bolometer (left) and the
    143-5 bolometer (right). Power spectra are averaged over 10\,000
    rings and are computed after removing the input simulated signal
    from the simulated data and using the estimated flags. The
    dot-dashed curve is the result without subtraction of long glitch
    tails, while the solid curve is with subtraction as done for the
    real data, and the dashed curve is for the same simulations but
    without glitches. Residual glitches are significantly reduced
    after template subtraction.}
  \label{fig:resnoisering}
\end{figure*}

For some representative bolometers we have evaluated the level of
contamination coming from glitches left in the data after processing at
the ring level, by comparing the power spectrum of the input simulated
noise projected on rings with the power spectrum of processed
simulated data, after removing the input sky signal from the
data. This is shown in Fig.~\ref{fig:resnoisering} for simulations of
data from the same two detectors as in the previous section: 143-2a,
containing a high rate of long glitches; and 143-5, with a low glitch
rate, but consequently more glitches below the threshold (see
Sect.~\ref{sec:interplg}), after averaging the spectra over 10\,000
rings. We also compare these two power spectra 
with the power spectra obtained without removing glitch
templates from the data, but using the same data flagging.

For bolometer 143-2a, we can see that without template subtraction the
glitch signal dominates over the noise on large scales,
$k<2\times10^{-3}$\,arcmin$^{-1}$, even after flagging.  We observe a
dramatic improvement in the noise power spectrum after removing long
glitch templates.  Nevertheless, glitch residuals still contribute to
the noise power at the level of 30\% at $k \lsim 2\times
10^{-3}$\,arcmin$^{-1}$ and less than 10\% at $k\gsim 6\times
10^{-3}$\,arcmin$^{-1}$.  We attribute the excess at low frequency to
errors in the subtraction of templates, which are expected to arise
due to uncertainties in the amplitude parameter determination after
$\chi^2$ minimization. Given the level of residuals in power spectra,
the impact of errors in template estimation is negligible at the percent
level, as seen in Sect.~\ref{templates}.  Undetected glitches
below the threshold contribute significantly at higher frequencies, at
the level of 5\% of the power for this bolometer, and cause the
observed excess over $0.03<k<0.1$ \,arcmin$^{-1}$. For bolometer 143-5,
the contamination by glitches before template subtraction below
$k<10^{-3}$\,arcmin$^{-1}$ is between 15 and 20\% and of the order of
8\% above $10^{-3}$\,arcmin$^{-1}$.  After glitch template
subtraction, the contamination reaches the level of 8\% for almost all
scales. This value corresponds to the level of contamination by
undetected glitches below the threshold, which is higher than for the
143-2a bolometer, in agreement with our modelling (see
Sect.~\ref{sec:interplg}).

This study shows that the remaining contamination from glitches is
below the instrumental noise level, even for the channels with the highest
glitch rates. Evaluation of the contamination at the map level and its effect
on the CMB power spectrum is
postponed until the release of the data for the full \Planck\
mission. Nevertheless, the contribution of glitches to the errors on
cosmological results \citep{planck2013-p11,planck2013-p08} has already
been accounted for, since the noise estimation is based on differences
of half--ring periods \citep{planck2013-p03}.

\section{Glitch characterization and interpretation}
\label{sec:characinterp}

In this section we focus on a more detailed description of  the
characteristics of the different glitch types and provide an
interpretation of their origin. This analysis is
complementary to the study based on ground tests presented in
\citet{particletest}.

\subsection{Evolution of the glitch rate over the mission}

\begin{figure}
  \centering
  \includegraphics[width=1.0\columnwidth]{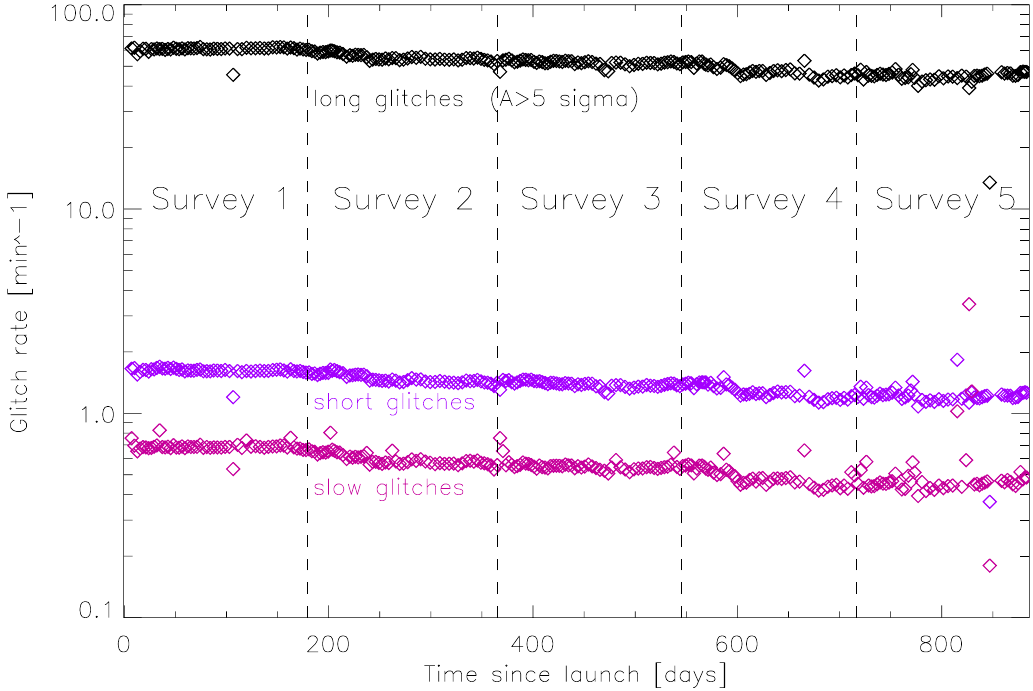}
  \caption{\label{rate_type_time} Mean rate of the three categories of
    glitches, i.e., long, short, and slow. These rates are computed
    using all 52 bolometers for long and short glitches, and the 16 PSB-a
    bolometers for slow glitches.}
\end{figure}

\begin{figure}
  \centering
  \includegraphics[width=1\columnwidth]{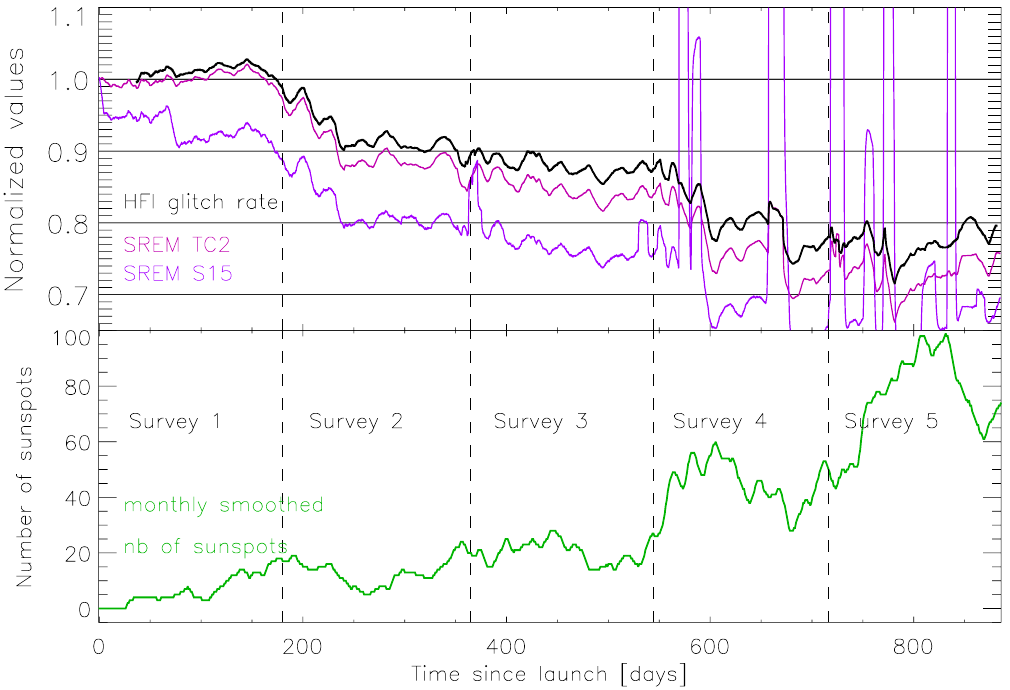}
  \caption{\label{TVar_glitch_Srem} {\it Top}: normalized glitch rate
    (black line), cosmic ray flux measured by the SREM for deposited energy
    $E > 3$\,MeV (purple), and $E > 0.085$\,MeV (pink), as a
    function of time. {\it Bottom}: monthly smoothed number of sunspots.}
\end{figure}

Over the course of the whole mission, the rates of each type of glitch
decreased (Fig.~\ref{rate_type_time}). This decrease is
universal for all bolometers (Fig.~\ref{rate_survey}). The
signal from diode sensor TC2, the most shielded of the three diodes in
the standard radiation event monitor \citep[SREM, ][]{SREM} located on
the Sun side of the \Planck\ spacecraft, follows a similar trend with
time (Fig.~\ref{TVar_glitch_Srem}).  This trend is expected
for Galactic cosmic rays modulated by the heliosphere of the Sun
\citep{gleeson68,bobik2012} and is observed by ground stations on
Earth and other spacecraft in the solar system
\citep{usokin11,wiedenbeck05,PAMELA_pHe_spec}. Taking together the
glitch rate, the SREM diode signal, and data from other studies, it is clear that the source
of glitches in the HFI bolometers is dominated by Galactic
cosmic rays.  Indeed, other sources, such as on-board radioactivity
and solar protons
\citep[as detected by the mirror on WMAP, ][]{jarosik2007},
were suspected to contribute to the glitch rate,
but would not follow the trends with time that are observed.

\subsection{Correlation in time}

\begin{figure*}
  \centering
  \includegraphics [width=1.0\columnwidth]{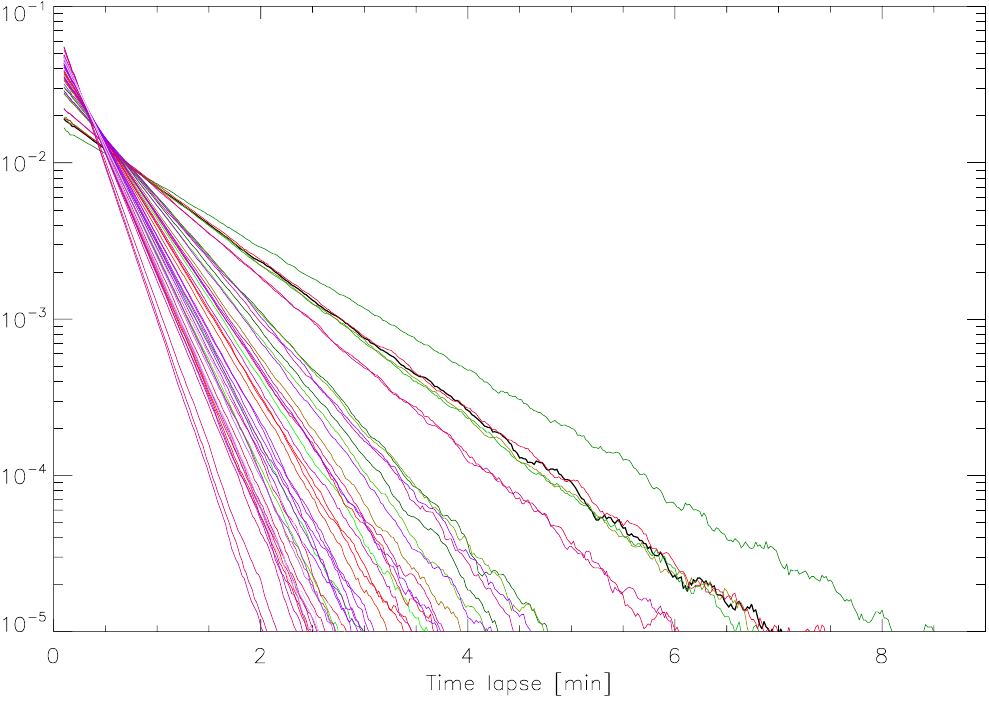}
  \includegraphics [width=1.0\columnwidth]{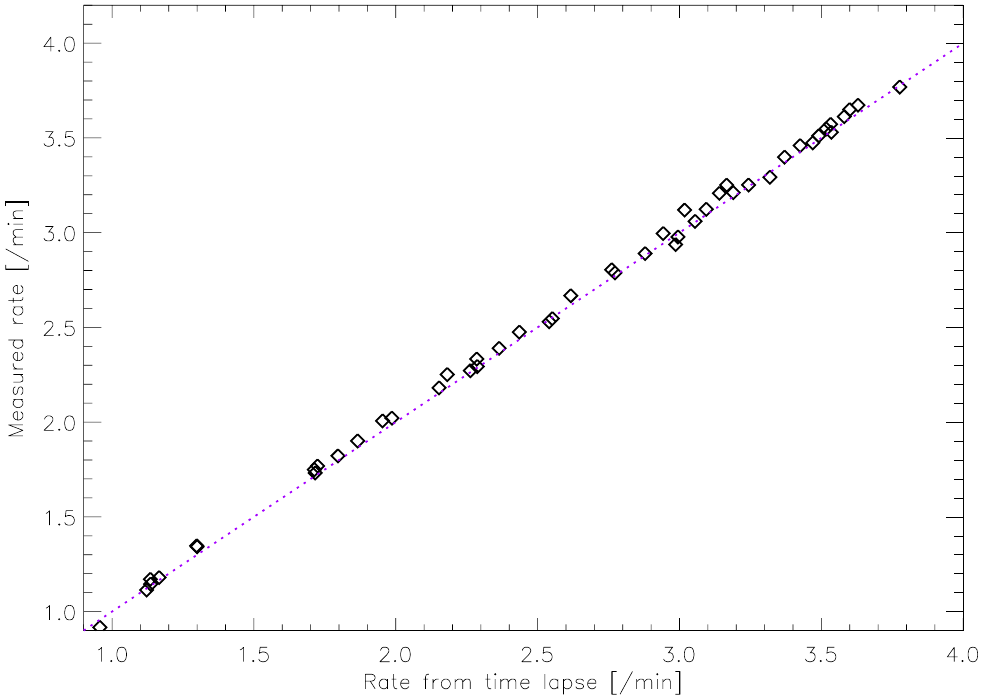}
  \caption{\label{poisson} {\it Left}: Time interval
    between two consecutive glitch events for each bolometer,
    normalized by the total number of events. Colours distinguish
    different detectors. We observe Poisson distributions, as expected
    for random events.  {\it Right}: Glitch rate derived from the
    slope of the time interval distribution, compared to the mission
    averaged actual rate for each bolometer.}
\end{figure*}

The time interval between two consecutive glitch events exhibits an
exponential distribution (left panel of Fig.~\ref{poisson}).  The
absolute value of the slope fitted between 1 and 2\,min is roughly
equal to the computed rate, simply obtained as the number of events
above 10$\,\sigma$ in six months (right panel of Fig.~\ref{poisson}).
Clearly, all the bolometer glitch distributions exhibit Poisson
behaviour.  These particular data are from Survey~3 when the glitch
rate was relatively constant and events above 10$\,\sigma$ were
selected to avoid any contamination by false events, solar flares, and
pile-up. We cannot check multiple simultaneous events because we
cannot distinguish between a pile-up of multiple events and a single
more energetic event. Still, the glitch data are compatible with pure
random events that are not correlated in time and this is consistent
with the cause of events being individual hits by Galactic cosmic ray
particles.

We also studied the coincidence of events between bolometers. We found
that the largest fraction of coincidences, around 99\%, are between
PSB-a and PSB-b detector pairs in the same module (see below).  The
remaining roughly 1\% of coincidences are particle shower events that
effect a large fraction of the detectors in the focal plane in
different modules (the shower events will be discussed in
Sect.~\ref{showers}).

\subsection{Interaction with the grid/thermistors}
\label{sec:interpsh}

There is clear evidence that the short events result from cosmic rays
hitting the grid or the thermistor. Indeed, these events have a fast
rise time and have a fast decay, and the transfer function built from
the short glitch template \citep[see][for a comparison]{particletest}
is in good agreement with the HFI optical transfer function
\citep{planck2013-p03c}, so the energy must be deposited in the
environment close to the thermistor.
\begin{figure}
  \centerline{\includegraphics[width=1.0\columnwidth]{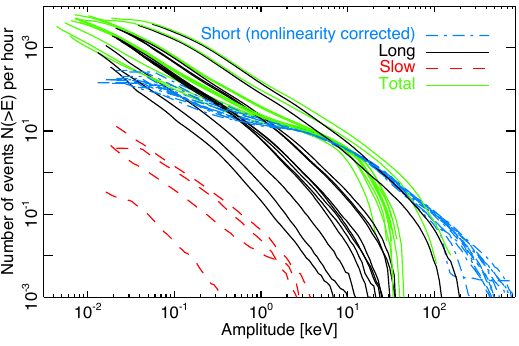}}
  \caption{Cumulative counts (per hour) for the three categories
    of glitches and for several bolometers: blue curves are for short
    glitches after nonlinearity correction above typically 10\,keV
    (as described in the text); black curves are for long glitches;
    red curves are for slow glitches; and green curves are for the
    total counts without the nonlinearity correction (which is why they
    lie below the blue curves). Glitches are not separated into the
    different categories below 10 times the rms noise (i.e., about
    20\,eV). All curves have been recalibrated in energy for each
    bolometer, such that the distributions of short glitches match at
    energies around 10\,keV (we use the same coefficient for all
    populations), the absolute calibration being fixed for one
    arbitrary detector. Note that the distributions of short glitch
    match very well, while there is a large scatter in the energy scaling
    for the other populations of glitches. }
  \label{fig:cumul}
\end{figure}
Figure~\ref{fig:cumul} shows the cumulative counts, i.e., the number
of events $N$($>$$E$) with an amplitude higher than a given value $E$
per hour, for short glitches, after converting the maximum amplitude
of each glitch to units of energy (keV) for one reference bolometer
(100-1b). The energy calibration for the other bolometers has been
scaled so that the cumulative distributions match the one for
the reference bolometer at energies of about 10\,keV. For this
conversion we use the measured heat capacity of the NTD crystal plus
grid system \citep[see][]{holmes2008},
since we assume that the energy of the events is
deposited in this system. The relative fitting of the energy is
necessary for the comparison between bolometers because of
uncertainties in the heat capacity of the system, and also in how much
the peak amplitude is reduced after averaging over the interval of one
sample within the electronics. Peak amplitude reduction is more
important for faster bolometers. The effect is significant for the
submillimetre channels, for which the time constant is very small. We
determined that the relative correction factor should be about
3 for 100--545\,GHz bolometers. We have also reconstructed
and corrected the counts for high amplitudes that are affected by
nonlinearity (typically above 10\,keV) by readjusting the values of
event energies using the measurements of the slow tails of short
events (which are not affected by nonlinearity since they are too low in
amplitude), instead of directly taking the peak amplitude. This allows
us to reconstruct the distribution more than a factor of ten above the
saturation level set by the electronics. We can see in
Fig.~\ref{fig:cumul} that the counts of short events for all detectors
match fairly well (considering the fact that no rescaling is performed
on the $y$-axis, but only in energy),
particularly around the energy bump.

Events in the bump are expected to be associated with cosmic rays
hitting the thermistor. Since the thermistors are identical for all
bolometers, we expect the same rates for all bolometers, which is
precisely what we observe (Fig.~\ref{fig:cumul}). Given the
dimensions of the thermistors of $30\,\mu{\rm m} \times 144\,\mu{\rm
  m} \times 341\,\mu{\rm m}$, and the fact that typical Galactic
protons of energies around 1\,GeV (the peak of the spectrum) deposit
minimum energies of $\approx 1.47\,{\rm MeV}\,{\rm cm}^2{\rm g}^{-1}$
in the Ge (which is the minimum of the stopping power for normal
incidence in the largest surface), we expect typical deposition
energies of 15\,keV in the thermistor (the first energy being
dominant, since it corresponds to the largest cross-section). This is
also in agreement with the energy measured experimentally, given
that the energy calibration is uncertain within about 50\%. The
amplitude of the bump in the cumulative distribution, which is at
about 10 events per hour, also matches with the expectations. Indeed,
by integrating the expected cosmic ray spectrum at L2, and considering
that low energy particles are absorbed by the spacecraft, we can
predict a total particle rate of $N \approx 5\,{\rm s}^{-1}{\rm
  cm}^{-2}$.  This leads to a glitch rate of about $8\,{\rm
  h}^{-1}$ on the thermistor.

The events depositing an energy of about 10\,eV to 1\,keV and
populating the power law seen in Fig.~\ref{fig:cumul} are expected
to be the result of cosmic rays hitting the grid. Calculations predict
a total rate of events on the grid from 7 to 90$\,{\rm
  h}^{-1}$, given that the surface area of the grid varies from
$4.77\times 10^5 \mu{\rm m}^2$ to $3.95\times 10^4 \mu{\rm m}^2$,
depending on the bolometer. We observe a higher number of events, by a
factor of 2--4, but (as we will see in the next paragraph) about half
of the events for PSB bolometers might result from electrons ejected
from the {\it other\/} grid by the impact of a cosmic ray proton. We
observe some dispersion in the amplitudes of the power-law
distributions from bolometer to bolometer, which is not obviously
correlated with the area of the grids. This is attributed to the
uncertainties in the intercalibration of energy to match the counts,
as detailed at the beginning of this section.

\begin{figure}
\centerline{\includegraphics[width=1.0\columnwidth]{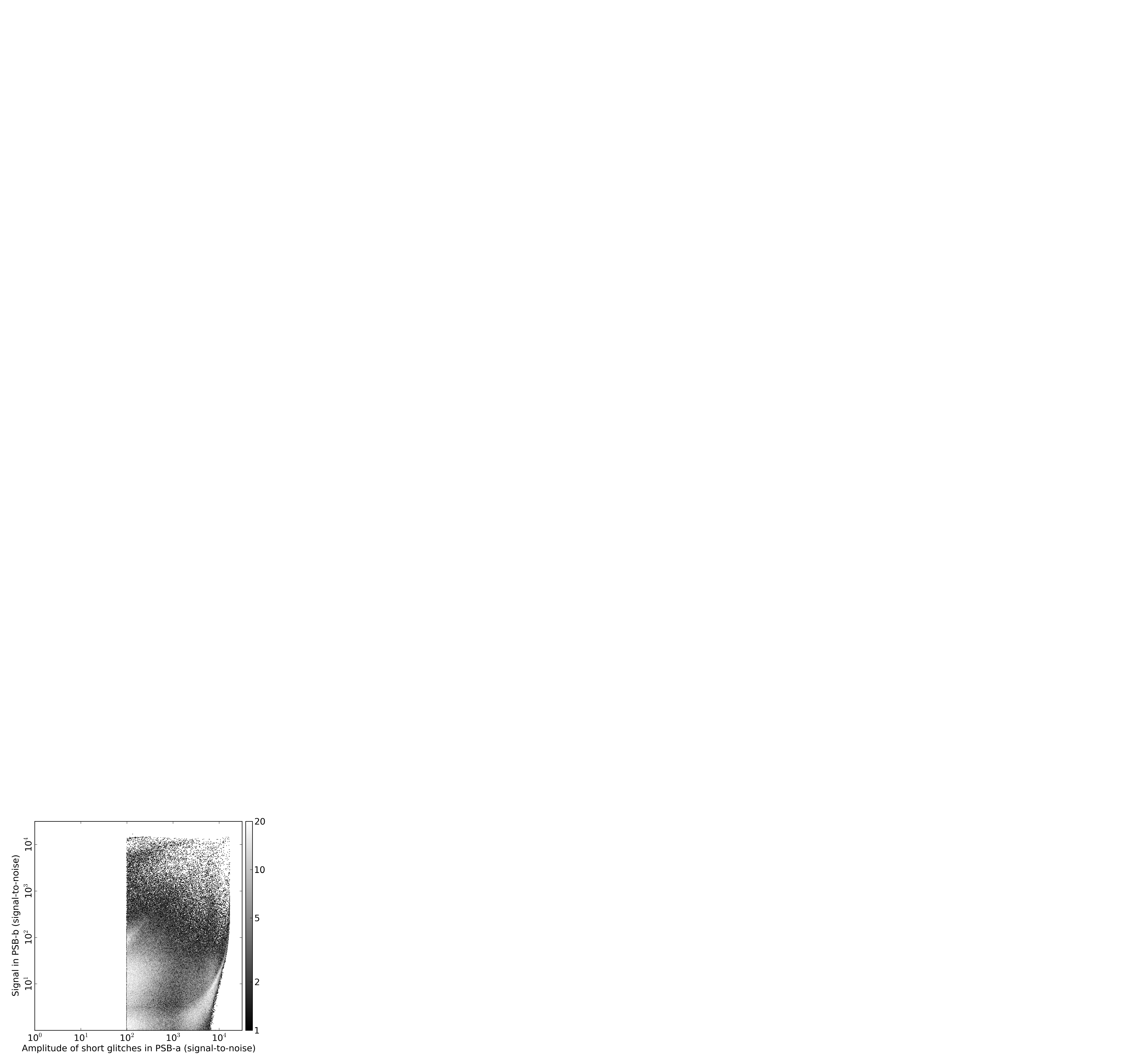}}
\caption{Representative short glitch coincidences in a PSB pair. The
  $x$-axis represents
  the amplitude relative to the noise
  of the detected glitch in the PSB-a bolometer, while the $y$-axis is
  the 
  value of the signal-to-noise ratio in the PSB-b bolometer. The
  grey-scale indicates the log of the number of events in logarithmic
  bins. Detected events are not shown below $100\,\sigma$ in PSB-a, as
  we expect a non-negligible contribution from long glitches, and the
  separation between long and short glitches is uncertain at low
  amplitudes. A large fraction of events are coincident between PSB-a
  and PSB-b.}
\label{shcoinc}
\end{figure}

\begin{figure}
  \centerline{\includegraphics[width=1.0\columnwidth]{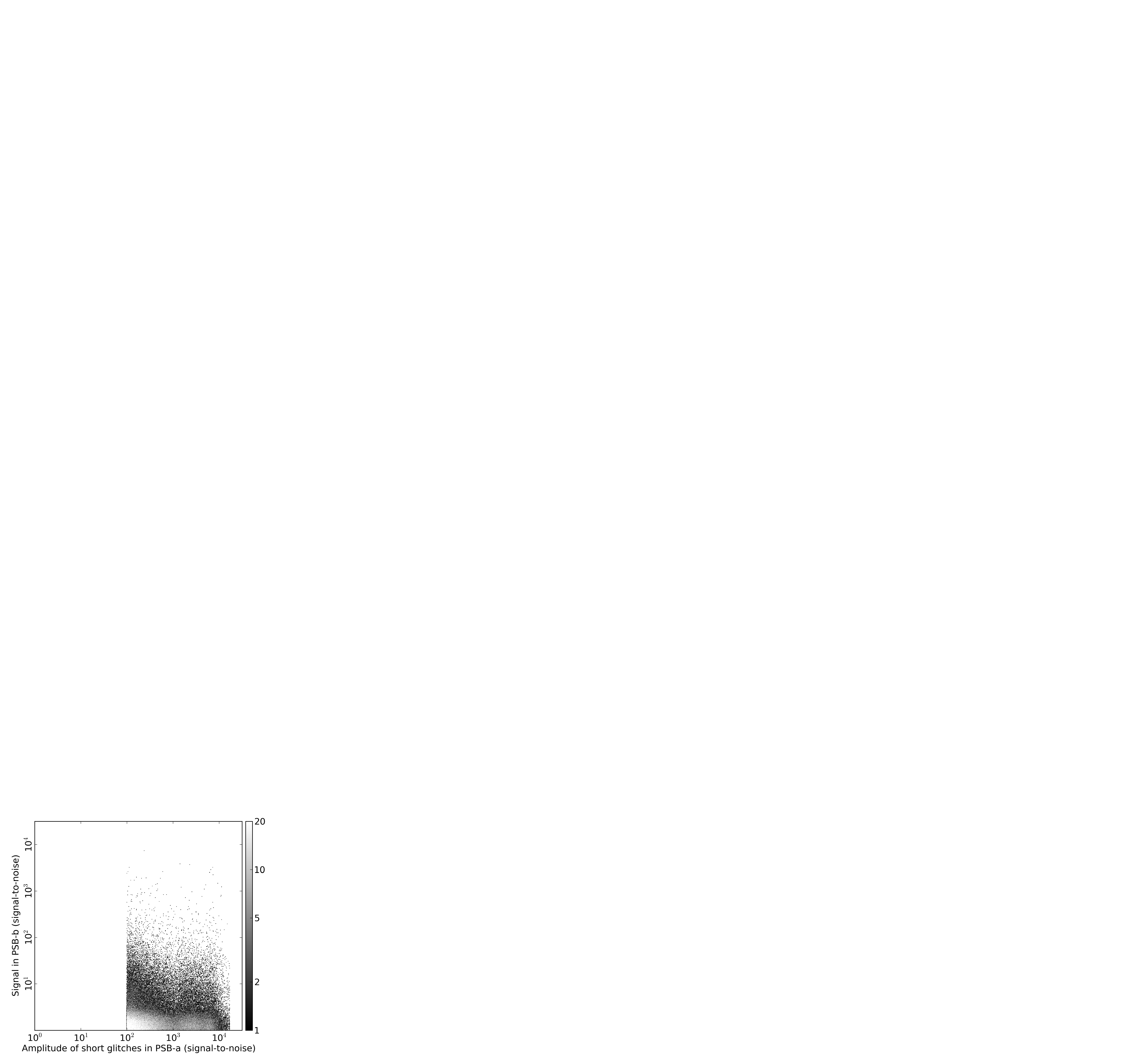}}
  \caption{Expected distribution for random coincidences, for comparison with
   Fig.~\ref{shcoinc}. The value of the
    signal-to-noise ratio in PSB-b for each short event in PSB-a is taken
    10\,000 samples afterwards. We use short events in PSB-a so that
    this distribution can be directly compared to the coincidence of
    short glitches.}
  \label{randcoinc}
\end{figure}

\begin{figure}
  \centerline{\includegraphics[width=1.0\columnwidth]{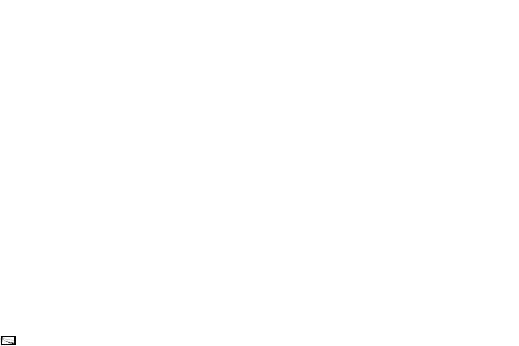}}
  \caption{Distributions of the signal values in PSB-b for short
    events with peak amplitudes between 300 and $1000\,\sigma$ in
    PSB-a (solid line). A random distribution of values in PSB-b is also
    shown (dot-dashed line). The excess of the solid curve over the dot-dashed
    curve for amplitudes above $3\,\sigma$ represents about 50\% of the
    total area. Thus, at least 50\% percent of the events are in
    coincidence between PSB-a and PSB-b.}
  \label{fig:coincEcut}
\end{figure}

We have performed a coincidence analysis of short events between PSB
pairs. At the arrival time of each short glitch detected in a PSB-a
bolometer, we have measured the level of signal in the PSB-b detector
in the same module. We compare the amplitude of the signal in the
PSB-b (without regard to the category of the signal in the PSB-b)
with the amplitude of the glitch at its maximum in the PSB-a, and
computed a 2-D histogram of the two quantities.  The 2-D distribution
of amplitudes for short glitches in the PSB-a is shown in
Fig.~\ref{shcoinc}.
The events below $100\,\sigma$ are not displayed since the separation
between short and long glitches is then uncertain. We observe that the
distribution is very different from the expected distribution for
random coincidence. The expected distribution of amplitudes for random
coincidences is evaluated by measuring the statistics of the signal in
the PSB-b at a time 10\,000 samples (about one minute, i.e., well after a short
glitch has terminated) after each event in the PSB-a, and this is
shown in Fig.~\ref{randcoinc}.
In particular, we can see a wide cloud of points in the 2D histogram
above about $10\,\sigma$ in PSB-b. A fraction of around 50\% of the
events populate the region of random coincidence. This is also shown
in Fig.~\ref{fig:coincEcut}, which displays the distribution of the
signal in the PSB-b for events in the PSB-a with amplitudes between
300 and $1000\,\sigma$. So there are strong indications that about
50\% of short glitches are seen in coincidence between PSB-a and
b. However, for a given amplitude of short glitch in PSB-a, we observe
a wide distribution of amplitudes in PSB-b. We have verified that
counterparts in PSB-b have no phase shift.

A possible model for these coincidences is that events hitting one of
the grids eject some electrons which hit the other grid. The rate of
50\% correlation can be explained by geometrical effects, such as
particles coming from the top or bottom; a particle coming from the
top could hit the PSB-a grid, extract electrons, and project those
onto the PSB-b grid, whereas a particle coming from the bottom and
hitting the PSB-a grid would not project electrons onto the PSB-b. The
observations indicate, then, that nearly all events hitting one of the
grids and coming from the appropriate direction must eject electrons
hitting the second grid and deposit about 10\% of the energy deposited
in the first grid. We also observe that some events clearly hit both
grids. These appear in the cloud of points (in Fig.~\ref{shcoinc})
with the same amplitude in both the PSB-a and b, representing about
2\% of the events. We observe a lower coincidence rate for events at
higher amplitudes corresponding to the bump. This is explained by the
hypothesis that those events correspond to direct impacts on the
thermistor, which is not aligned with the grid of the other bolometer.
Nevertheless, we have seen that {\it all} those high energy events are
in coincidence with a small amplitude event in the second bolometer.
This is attributed to the cross-talk between bolometers in a pair, as
discussed in Sect.~\ref{cross-talk}. The nonlinearity, appearing as
the curved shape in the coincidences, can be entirely explained by the
saturation of the highest amplitude glitches. Nevertheless, we can
evaluate the level of cross-talk to be between 0.01 and 0.44\%,
depending on the bolometer pair, for about $3000\,\sigma$ events (for
which the effect of saturation is small). The high amplitude events
correspond to an energy of $\approx$ 15\,keV left on the
grid+thermistor by a particle; this is very close to the expected
energy deposited by a 1\,GeV proton onto the thermistor.

\subsection{Interaction with the wafer}
\label{sec:interplg}

\begin{figure}
  \centerline{\includegraphics[width=1.0\columnwidth]{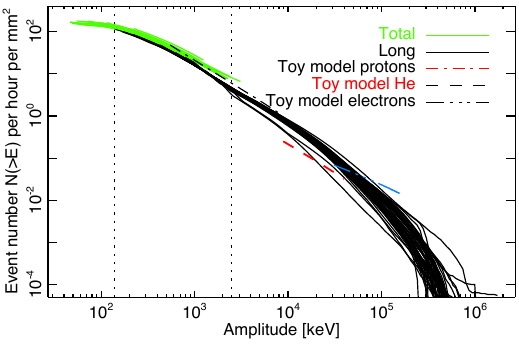}}
  \caption{Cumulative distribution $N$($>$$E$) of long glitches per
    unit surface area of the wafer for all bolometers used for the
    scientific analysis. The black solid lines correspond to selected
    long glitches. The green lines correspond to total glitches, but
    are representative of long glitches, since they dominate at low
    energy. We have used a relative calibration of the energy
    (indicated on the $x$-axis) such that the cumulative counts for
    all bolometers match at $10^3$\,keV. The absolute calibration of
    the energy is set so that the faint-end break of the counts
    matches the expected deposited energy of about 140\,keV in the Si
    die. Nevertheless, we observe that the shape of the cumulative
    counts match for all detectors, and with a low dispersion. The
    break in the counts at the faint end is very significantly
    detected; this is our best indication from flight data that long
    glitches result from particles hitting the Si die of the
    bolometer. Dot-dashed and dashed lines correspond to the predicted
    power-law distributions of deposited energy from protons and
    helium, respectively, with a toy model for the interaction on the
    wafer using the incident particle spectrum at L2 for energies
    significantly higher than the stopping power energies. Predicted
    minimum deposited energies for He and for protons are indicated
    with dotted lines. We interpret the second apparent bump (seen as
    a slight break in the spectrum at around 20 times the energy of
    the main break) as the signature of He nuclei.}
  \label{fig:cumulLg}
\end{figure}

We identify the long glitches as being produced by cosmic rays hitting
the silicon die. This was first indicated by the ground tests
\citep{particletest}, showing that the NTD thermometer is sensitive to
a temperature change of the Si die. The HFI ground-based calibration
data show a rate of events compatible with the cosmic ray flux at sea
level over the Si die surface, and the ground-based data also show
that almost all these events are in coincidence between PSB-a and
PSB-b. Our understanding is the following. Hot carriers are generated
by the particle impacts on the Si die. These rapidly decay into hot
ballistic phonons which cause the rapid (athermal) rise in NTD Ge
temperature. The bolometer temperature then decays with fast optical
time constant. The next slower time constant is due to the thermal
response of the Si die \citep{particletest}. The slowest thermal
response is due to the heat flow through the copper module and into
the 100\,mK heat sink (Spencer et al. in prep.)
This hypothesis is reinforced by the comparison of cumulative counts
$N$($>$$E$) of long glitches from bolometer to bolometer (Fig.~\ref{fig:cumulLg}).
To make this comparison, we have
normalized the counts so that we are evaluating the number of events
per unit time and per unit surface area of the Si wafer (because
the total surface varies from bolometer to bolometer). Also, as for
short glitches, we have intercalibrated the energies of events between
bolometers by matching the different counts at measured energies
around $0.05$\,keV, which corresponds to a deposited energy on the Si
die of about $10^3$\,keV after absolute calibration.  This absolute
calibration is performed so that the observed faint-end break of the
counts, which is clearly visible in the figure, matches the expected
minimum energy deposited by GeV protons on the Si die (which is about
140\,keV, for normal incidence, as described later).
This calibration of the model is necessary, since we only measure
the energy in the grid+thermistor system, and the heat capacity and
link conductivities are not known with sufficient accuracy.
Additionally, we use the values of the peaks to record the amplitudes
of glitches that are not representative of thermal processes, since
the fast part of long glitches is expected to result from ballistic
phonon effects.

Figure~\ref{fig:cumulLg} shows the cumulative counts for all the HFI
bolometers used for the scientific analysis.  We clearly see that they
all match in shape and amplitude, with very small scatter. For
energies below 20 times the rms noise, we have also computed the
cumulative counts for all events, without distinction between families
(green curves in the figure); we have already seen that long events
dominate at low energy, so the total counts are representative of the
counts of long glitches. The faint-end break in the counts is detected
without ambiguity. This is an important result, since it shows that
there are a limited number of undetected low energy glitches in data,
which might otherwise affect the cosmological results. From this
limit, we measure that the total rate of events penetrating the
shielding of the satellite around the bolometers is about $4.5\,{\rm
  s}^{-1}{\rm cm}^{-2}$. This is in good agreement with the
expected value in the bolometer environment computed in
Sect.~\ref{sec:interpsh} of $5\,{\rm s}^{-1}{\rm cm}^{-2}$.  The slope
and amplitude of the distribution can be predicted using a simple
power-law model of the interaction of primary Galactic protons and He
nuclei with the Si die, as shown in Fig.~\ref{fig:cumulLg}. We observe
a good match of the model with the data, showing that particles
detected by \Planck\ at the energies above the faint-end break are
primary Galactic protons. The barely apparent second bump in
cumulative counts at deposited energies around 3000\,keV may
correspond to the contribution from He nuclei, which are expected to
contribute about 10\% of the counts at those energies. The fact that
we observe an excess in the counts at the expected energy of the
minimum stopping power for He reinforces our hypothesis for the origin
of long glitches. High energy Galactic electrons could also contribute
to the measured counts for high deposited energies.

The large scatter in the distributions of long glitches with
respect to different detectors, after calibrating the energy on the short
distributions, is shown in Fig.~\ref{fig:cumul}.  This is due to the
variation of thermal links between the grid and the Si die, as well as
variations in the Si die heat capacity, reducing or increasing the fraction
of energy transmitted to the thermistor for both phonons and thermal
processes. The detectors with smaller glitch counts (e.g., 143-1a and
143-5 in Fig.~\ref{rate_type_time}) are those with a smaller fraction
of the energy propagating from the wafer to the grid. For those
detectors we do not see the break in the counts, since it should be below
the detection threshold. We observe that in general PSB-b detectors are
more sensitive to long glitches than PSB-a detectors (although a few
exceptions exist).

Detectors with smaller rates of glitches should have more undetected
glitches than detectors with a higher rate. This hypothesis is
verified with the help of the cross-correlation analysis between
bolometer signals from the same PSB pair, described in
Sect.~\ref{methodres} (see Fig.~\ref{crossspect} in particular).  We
observed that bolometer pairs with smaller glitch rates are those with
higher correlations around 1\,Hz, which is attributed to glitches
below the threshold.  On the other hand, pairs for which one of the
bolometer has a visible faint-end break in the counts have smaller
correlations in the noise.  The impact of this on final results after
processing is studied in Sect.~\ref{errorestim}.

\begin{figure}
  \centerline{\includegraphics[width=1.0\columnwidth]{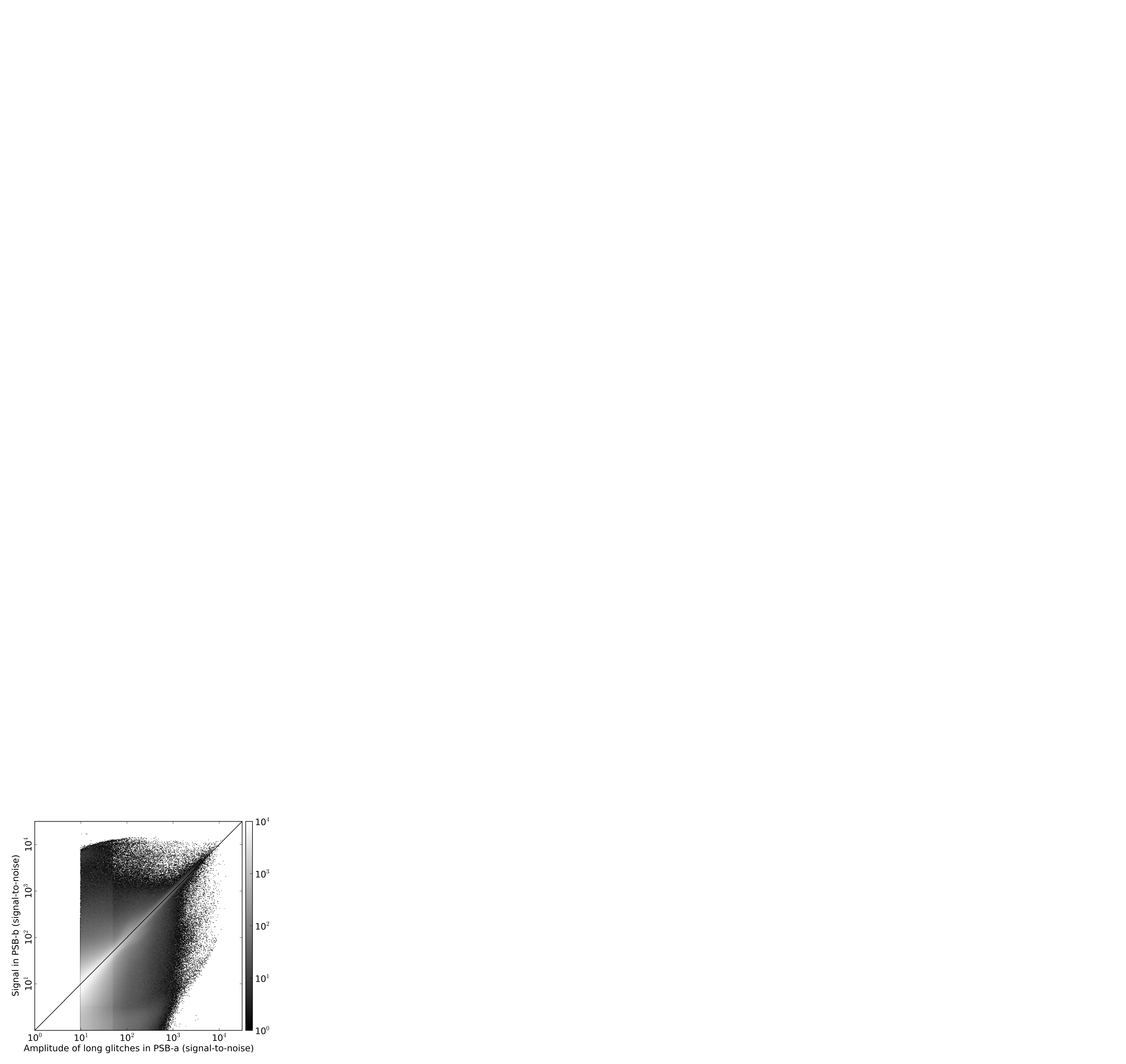}}
  \caption{Representative long glitch coincidences in a PSB pair. See
    description in Fig.~\ref{shcoinc}. Events above $50\,\sigma$
    are selected if they match the long glitch
    template. All events (without selections) are included below $50\,\sigma$,
    since the long glitch population dominates at low amplitudes. The
    solid line indicates identical signal in both the PSB-a and
    PSB-b. Most (if not all) long events in PSB-a are associated with
    significant events in PSB-b, and a large fraction have
    approximately the same amplitude in PSB-a and PSB-b.}
  \label{lgcoinc}
\end{figure}

We have already seen that events hitting the bolometer Si die must
deposit a fraction of their energy, which is typically the minimum
stopping power of the Si for events close to normal incidence and with
energies of 140\,keV. Then, most of the events must instantaneously
hit the two wafers in PSB pairs. This is compatible with what we
observe by measuring the coincidence of long events in PSB-a with
signal in PSB-b, shown in Fig.~\ref{lgcoinc},
as we did for short events.
Indeed, we find that between a third and a half of the events deposit
the same energy in both PSB-a and PSB-b detectors, with some intrinsic
dispersion, as most of the points lie along a line in the figure. The
factor between amplitudes of coincident events is almost 1 in
signal-to-noise units for this PSB pair, but can vary between 0.25 and
4, depending on the bolometer pair. This factor becomes 1 after
recalibrating in units of deposition energy on the Si die, as
expected. Furthermore, 100\% of the events (at least for amplitudes
higher that $300\,\sigma$) happen in coincidence between the two
bolometer pairs, but with different energy deposition, by a factor of
300 at maximum. We cannot state this for events below 300\,$\sigma$,
since a significant fraction of the counterparts in PSB-b are below the
noise level.  We could explain the difference in energy deposition
between PSB-a and PSB-b by the fact that some relatively low energy
events lose a significant fraction of their energy in the first Si die
and then deposit more in the second, since the stopping power is a
decreasing function of the energy for the considered particles.  Some
electrons could also be ejected from the first wafer and interact with
the second wafer (as discussed for the grid interaction), but the
associated signal would be diluted in the signal from the primary
particle, since the events are in coincidence in the two wafers.  In
principle we should have a small proportion of events hitting the
corner of the bolometer wafer, with no counterparts in other
bolometers, but we do not have evidence for such events.  The number
of events in the distribution along the line of one-to-one correlation
is of the order of 50\% of the total number for events with amplitudes
lower than about $300\,\sigma$ or  above about
$1500\,\sigma$, and is around 35\% for amplitudes in between.  A
complete modelling of the interaction of particles with the two Si
dies, including accurate physical modelling of the interaction, is
postponed to a forthcoming publication.

Crucially, the coincidence study indicates that the contribution to
the glitches by secondary particles, e.g., ``delta'' (or secondary) electrons,
is negligible, since those low energy events would deposit all their
energy in the first wafer without a counterpart in the second.

\begin{figure}
  \centerline{\includegraphics[width=1.0\columnwidth]{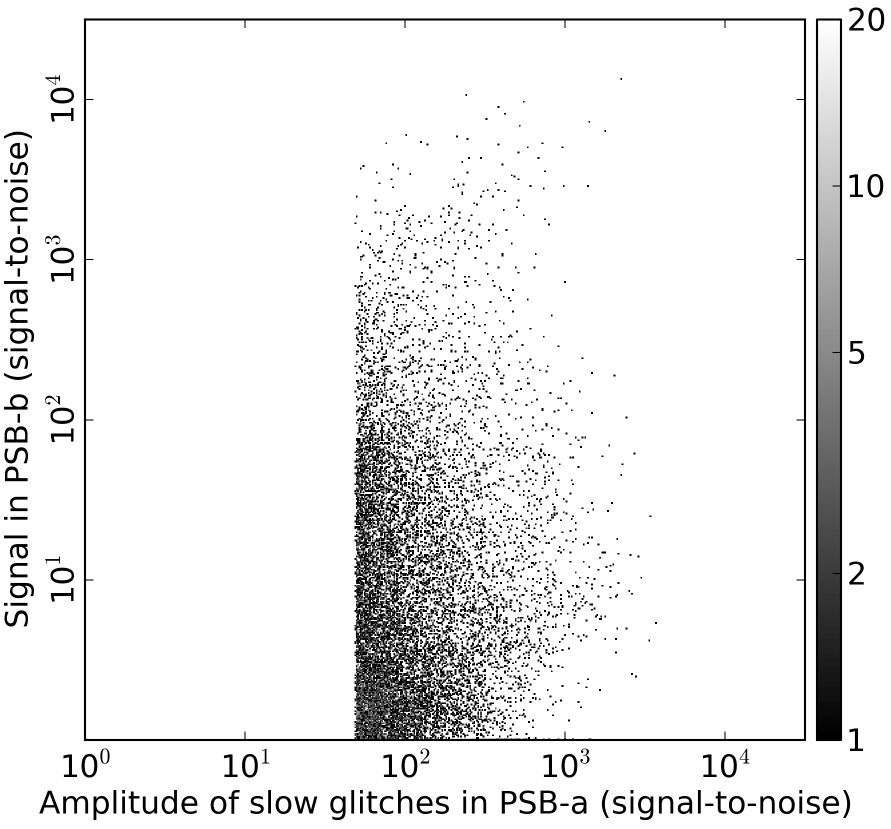}}
  \caption{Representative slow glitch coincidences in a PSB pair. See
    description in Fig.~\ref{shcoinc}. No event is shown
    below $50\,\sigma$, to avoid leakage from the long population. All
    slow glitches in PSB-a above about $1000\,\sigma$ are associated
    with significant events in PSB-b. Below about $1000\,\sigma$
    the majority of events are associated with significant events in
    PSB-b.  For the remaining fraction the fact that we do not see a
    significant signal might be simply a threshold effect.}
  \label{sncoinc}
\end{figure}

We have also performed a coincidence analysis at the location of slow
events in PSB-a. Distributions are shown in Fig.~\ref{sncoinc}.
As already stated, slow glitches are seen only in PSB-a. Nevertheless,
there is a significant signal in PSB-b associated with slow glitches
in PSB-a. Indeed, we can see that the distribution of counterparts in
PSB-b is very different from the random distribution shown in
Fig.~\ref{randcoinc}, which means that slow events in PSB-a are
associated with signals in PSB-b. The counterparts generally have a
wide range of amplitudes for a given energy of slow glitches, which
makes it difficult to measure the associated time constants. Visual
inspection of those events allows us to distinguish a variety of
counterparts: some have faster decay than slow glitches and have
similar rise times of around 10\,ms; and some others (the larger
energy ones) are similar to long glitches.

\subsection{Cross-talk between bolometers}
\label{cross-talk}

A very high amplitude short event in one of the bolometers of a PSB
pair is always seen in coincidence with a small event in the other
bolometer. This is attributed to cross-talk between bolometers. The
estimated level is about $10^{-3}$, as can be seen in
Fig.~\ref{shcoinc} for a particular pair; the level varies
between 0.01 and 0.44\% for different pairs. These events show a
phase shift, which varies from 0.25 to 1.4 samples, depending on the
pair, and the transfer function of the cross-talk signal is different
from the primary glitch transfer function. We observe longer
time constants for the cross-talk signal than for the bolometers,
ranging from 10 to 30\,ms. The origin of cross-talk between detectors
from the same PSB pair might be electrical or thermal; there is no
clear evidence at this point.

We have also attempted to estimate the cross-talk signal between bolometers
which are not mounted in the same pair. To do so we have averaged the
bolometer signal at locations of detected high energy events in one of
the bolometers. We found a significant (but very small) cross-talk
signal of the order of $3\times 10^{-5}$ for some pairs of detectors
within the same electronics belt. However, we did not measure any
significant signals in data from other bolometers, with an estimated
limit of $3 \times 10^{-6}$.

\subsection{Solar flares}

\begin{figure}
  \centering
  \includegraphics[width=1.0\columnwidth]{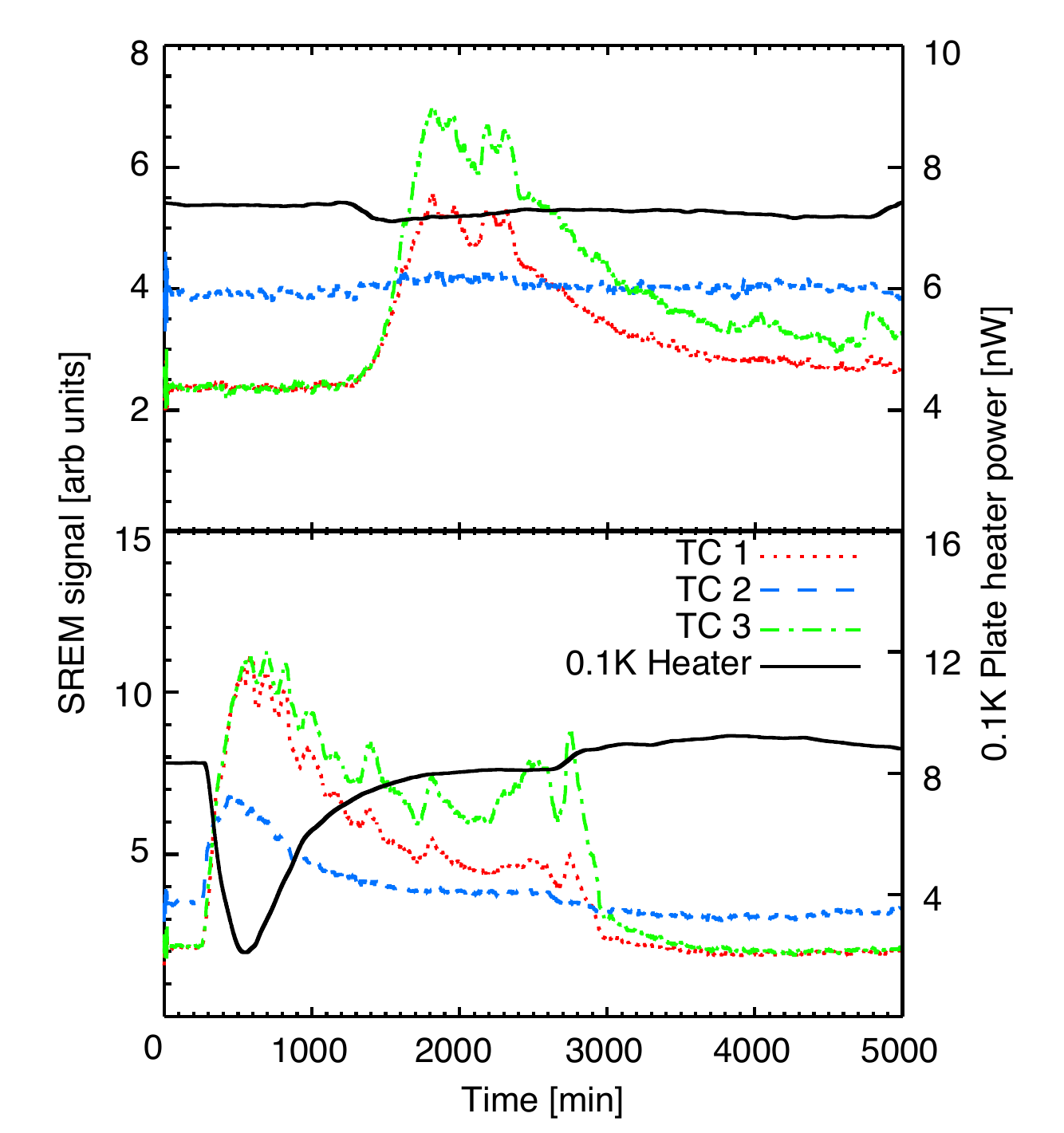}
  \caption{\label{flare_PID}Output of SREM diodes TC1, TC2 and TC3
    (left axis) and temperature control heater power on the 0.1\,K plate
    (right axis), and as a function of time for large solar flares on
    7 March 2011 (top) and 4 August 2011 (bottom).}
\end{figure}

For most of the mission, solar activity was remarkably low
\citep{mewaldt10}. In 2011, for the first time in the mission, there
were several large solar flares.  These flares provided useful test cases
for correlating the signal measured on the outside of the
spacecraft (with the SREM) with signals due to particle impacts on HFI.
The glitch rate noticeably increases during solar flares. The heater
power used to regulate the 0.1\,K plate decreases with increasing
signal, and hence the particle flux, measured by the SREM diode sensors
(Fig.~\ref{flare_PID}).  The timescale of each
flare was slow enough that the temperature control loop was able to
compensate for the bulk heating produced by the increased particle
flux, as shown by the small phase shift of the heater power
compared to the onset of the signal in the SREM.
We see in Fig.~\ref{flare_PID} that the peak measured signal for TC1
and TC3 differ by only about 40\% for the two flares. The signals of
the heater power and TC2 are similar to each other in each
flare. However, the peak signal of each is very different when
comparing the two flares. In addition, there is structure in the
signal for TC1 and TC3, which is not seen in TC2 or the heater power
response.  We find that this correlation holds between the heater
power and TC2 for all flares.  The diode TC2 has the most shielding of
the three diodes in the SREM, 1.7\,mm of aluminium and 0.7\,mm of
tantalum, which passes only ions and protons with energies
$>39$\,MeV. The other diodes are shielded by 0.7\,mm of Al for TC3 and
1.7\,mm of Al for TC1.  This demonstrates that the spacecraft and the
instrument surrounding the bolometers act to shield particles with
energies at least up to 39\,MeV, as well as all solar electrons.  This
is similar to the stopping power of about 1.5\,cm of
Al.\footnote{\url{http://physics.nist.gov/Star}}.

\subsection{High coincidence events}
\label{showers}
 
Since the beginning of the \Planck\ mission, large temperature increases
of the 0.1\,K plate have been observed through the bolometers and
thermometers. These events have two specific characteristics. There are
a high number of glitches detected at the same time in multiple
bolometers, and there is an increase of the focal plane temperature in the
range of nanokelvins to microkelvins.  In this section we will analyse the characteristics of
these high coincidence events (HCEs).

\subsubsection{Detection}

HCEs are identified by detecting a large number of precursor glitches
in coincidence in different detectors (so called ``touched'' bolometers)
on the 0.1\,K plate. These coincidences happen before a
temperature excursion of the 0.1\,K plate. Other automated techniques
that focus on finding only the temperature excursion were applied to
the data, but were not successful at identifying HCEs in the time
stream. One reason for this is that for most of the events this
temperature rise is below the thermometry noise level.

\begin{figure}
\centering	
\includegraphics[width=1.\columnwidth]{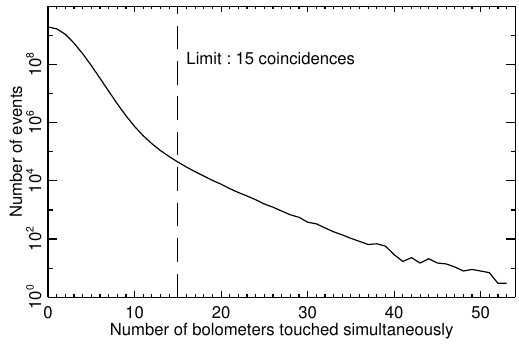}
\caption{Coincidence histogram (solid line) processed for the entire
  mission, based on the result of the glitch detection
  algorithm. Random coincidences dominate for low numbers of ``touched''
  bolometers.  Coincidences above 15 bolometers are real
  correlations (dashed line).}
\label{glitch_histo}
\end{figure}

Figure~\ref{glitch_histo} shows the histogram of the number of events
in coincidence for the full mission for 15\,ms bins. The distribution
at low number of coincidences is nearly compatible with random
coincidence, given the measured rate of glitches per bolometer. The
distribution deviates from random coincidences above 13 ``touched'' bolometers.
We set the detection threshold at 15 in order to select only
real coincidences linked with HCEs.
We detect around 100\,000 HCEs for a theshold
of 15, or an average of 5 events per hour.

\subsubsection{Different types of HCE}

\begin{figure*}
\includegraphics[width=1.\columnwidth]{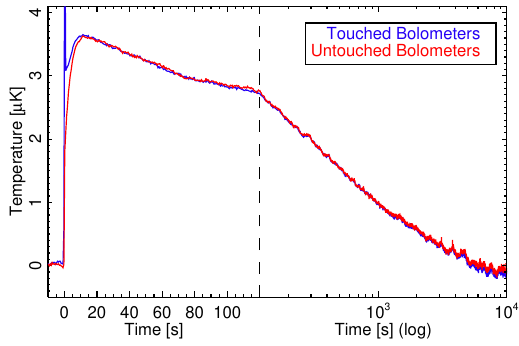}
\includegraphics[width=1.\columnwidth]{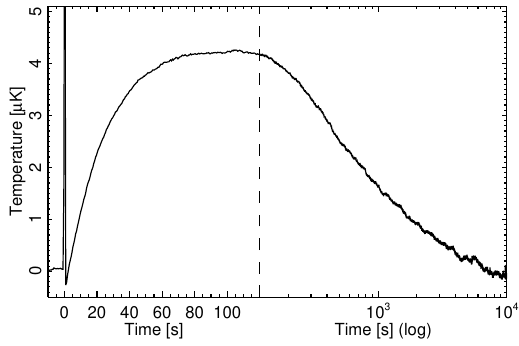}
\caption{Stacking of 69 fast HCEs for bolometers detected as
  ``touched'' and ``untouched'' and 72 slow HCEs for all
  bolometers. The scale changes from linear to logarithmic at
  100\,s. For fast HCEs (\textit{left panel}), the limit is clear, and
  the stacking of ``untouched'' bolometers gives the thermal time
  constant of the copper/stainless steel bolometer plate heating. For
  slow HCEs (\textit{right panel}) we observe a temperature decrement
  lower than 1\,$\mu$K for about 1\,s before the temperature rises,
  and this is not understood.}
\label{shortlongstack_ele}
\end{figure*}

For each event above the thermometry noise level we calculate
separately the median of the temperature time stream for ``touched''
bolometers and for bolometers without a precursor glitch, or
``untouched'' bolometers.  We can separate events into two categories,
fast and slow, which differ in the rise time of the temperature
(Fig.~\ref{shortlongstack_ele}).  The maximum temperature value is calculated 
in the range 8--12\,s after concidence for fast events, or 80--120\,s after the coincidence for
slow events.

For fast HCEs, coincident glitches occur in several (but not all)
bolometers, and are grouped together on the 0.1\,K plate. The
temperature rise time constant is around 5\,s. The decay time
constant is approximately 20\,min and matches the thermal time
constant of the 0.1\,K plate.

\begin{figure*}
\includegraphics[width=1.\columnwidth]{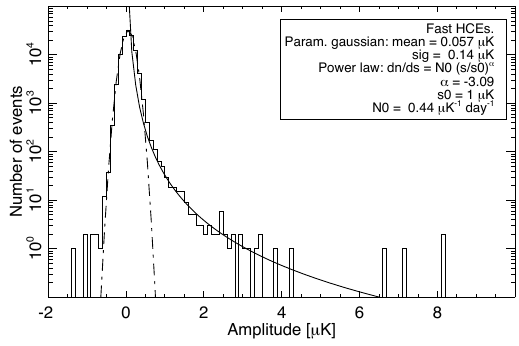}
\includegraphics[width=1.\columnwidth]{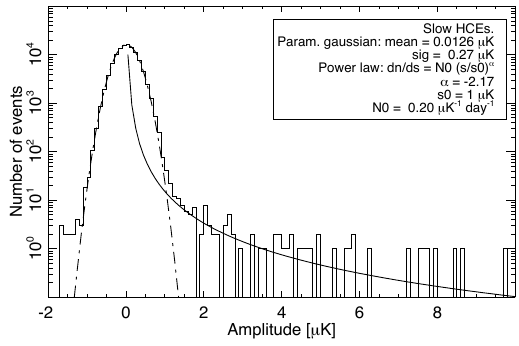}
\caption{Amplitudes of fast (\textit{left panel}) and slow (\textit{right
  panel)} HCEs. Amplitudes are computed by jointly fitting fast and
  slow templates. Solid lines correspond to power laws fitted to the
  tails of distributions (above 1\,$\mu$K), the high amplitude fast
  events above 7\,$\mu$K being removed from the fit; the number of
  those high amplitude fast events is not compatible with the power-law
  distribution. Dot-dashed lines are Gaussians fit to the
  distributions near zero amplitude and are representative of fitting
  errors. Parameters of the fitted distributions are given in the
  figure legend. Fast events (and to some extent slow HCEs) have a
  varying temporal shape from event to event. It appears that the
  variation can be fit by a contribution from the slow HCE
  template. This explains in part the negative tail of the
  distributions of fitted amplitudes. The rest may be
  confusion between events inducing an anti-correlation of amplitude
  parameters.}
\label{Histo_ampl}
\end{figure*}

The precursor glitches for a fast HCE are short or long glitches. The
separated ``touched'' and ``untouched'' bolometer temperature
time stream stacking averages, for fast HCEs greater than 1.5\,$\mu$K,
are shown in Fig.~\ref{shortlongstack_ele}. The ``untouched'' curve
rises slowly, but decays in phase with the ``touched'' one after all
bolometers thermalize with the entire 0.1\,K plate. The distribution
of fitted temperature amplitudes for all the fast HCEs (coincidence of
greater than 15 bolometers) obtained by template fitting is given in
the left panel of Fig.~\ref{Histo_ampl}. We observe that fast events
above 0.8\,$\mu$K are individually detected. For events detected with
signal-to-noise ratio less than one, we find a non-zero amplitude of
$0.057\,\mu$K. We also observed that the rate of fast HCEs is
modulated in the same way as cosmic rays over the mission.

The fact that the ``touched'' bolometers are grouped in the focal
plane and precursor glitches of fast HCEs are the same as for
individual particle hits directly on the Si die (for long glitches) or
grid (for short glitches), leads us to conclude that these events are
due to showers of secondary particles over part of the focal plane.
The temperature increase of the 0.1\,K plate results from the low
energy secondary particles from the shower, which are stopped by the
bolometer plate. The rate of the most energetic events is consistent
with the geometry of the 0.1\,K plate and the flux of Galactic cosmic
rays that are energetic enough ($\gg$\,GeV) to produce such showers
\citep{PAMELA_pHe_spec}.

For slow HCEs, the coincident precursor glitches occur in all, or
nearly all, bolometers. As shown in Fig.~\ref{shortlongstack_ele},
there is a small temperature decrease around 1\,s after the
precursor glitch and before the temperature rise, with a time constant
of about 30\,s and, as observed for fast HCEs, a decay time constant
of about 20\,min. This initial decrease prior to the rise is not
yet understood.

The precursor glitches for these HCEs are very specific and seem to be
different from any of the categories studied in
Sect.~\ref{sec:characinterp}. They have a rise time of about 10\,ms
and a decay time that varies from event to event, but is of order
100\,ms, which is a lower value than for the slow glitches
(see Sect.~\ref{slowgl}).  The distribution of fitted temperature
amplitudes for all the slow HCEs is shown in Fig.~\ref{Histo_ampl}
(right panel) and the validation threshold value is 1.3\,$\mu$K. Slow
HCEs have a less steep distribution than fast HCEs, and dominate above
2\,$\mu$K. In contrast to fast HCEs, we have observed that the rate of
slow events decreases through the mission.  We use the measured thermal
properties of the 0.1\,K plate \citep{planck2011-1.3} to estimate that slow
HCEs correspond to an energy deposition of around 1\,TeV, with a rate
that is not consistent with the Galactic cosmic ray
spectrum \citep{PAMELA_pHe_spec}.  We have also considered elastic
relaxation of cracks and anomalous response of the 0.1\,K stage PID
loop as possible explanations.  However, we have yet to reproduce a
slow HCE experimentally on the flight focal plane, for example with
temperature or heater steps, or in analogous ground tests.  At this
time, we have not identified a physical cause for the slow HCE events.

\subsection{Particle interaction with bolometer plate}

It has been shown in \cite{planck2011-1.3} that interactions of
particles with the bolometer plate are responsible for the correlated
low frequency noise between bolometers (see Fig.~\ref{crossspect} and
caption). This excess noise was not seen in ground measurements
\citep{pajot2010} and is modulated with solar activity in the same way
as the glitch rates.

Although the mean temperatures of the 0.1\,K and dilution cooler
plates are stabilized with great precision, this is only at a single
point near the control heater.  Distributed heating from particles and
other sources causes gradients within the plate. As described in
\cite{planck2011-1.7}, the signal from the two dark bolometers was
used to track the temperature of the 0.1\,K plate, independent of the
control thermometer.  NTD thermometers could not be used to control
the temperature variations because of the very hit glitch rates due to
their large volume. The signal from each dark bolometer was smoothed
over a running window of 2\,min (or about two full rotations of the
spacecraft) and combined to form a thermal template, which was then
cross-correlated with each bolometer and used to remove long-term
drifts from the bolometer time streams. Both the control heater power
and the thermal templates are correlated with diode sensors in the
SREM, indicating that a significant source of heating (although small,
$<10\,$nW) comes from interaction with the particle flux.

We have evaluated the contribution of fast HCEs to the bolometer noise
power spectrum. We start by assuming that the distribution of the
amplitudes of fast events follows the fitted power law shown in
Fig.~\ref{Histo_ampl}. Since the power law has a high index (around
3), some break has to occur towards the lowest amplitudes.  By
choosing a limit about 10 times lower than the amplitude corresponding
to where the total counts match the detected number of coincidences
for 15 ``touched'' bolometers, we found that fast HCEs might account
for approximately 10\% of the correlated low frequency noise between
bolometers.  By selecting events above 15 coincidences only, we are
not counting all actual coincidences, since coincidence of a few
bolometers ($<10$) are overwhelmed by the random events.  We have
tried techniques such as fitting HCE templates over a sliding window
in the temperature of the 0.1\,K plate (measured by the thermal
template), but have not found one that successfully detects the bulk
thermal response of the 0.1\,K plate from fast HCEs without the
precursor coincidence flagging.  However, even a modest increase in
the total HCE events with $<10$ bolometers in coincidence could
explain all of the correlated noise. We are pursuing models and
analysis techniques to quantify more accurately the number and effect
of these undetected coincidences.

\subsection{Other events}
\label{extraevents}

The origin of slow glitches is not entirely understood.
Laboratory tests using alpha particles were unsuccessful at reproducing
slow glitches. Nevertheless, the similarities with long glitches
suggest that both populations have the same physical origin, and that
they correspond to some energy injected into the wafer.  A key feature
of the slow glitches is that they only occur in the PSB-a detectors. The
mechanical mounting of the PSB-a bolometers is different from that of the
PSB-b and the SWB bolometers. Tests using heaters and
thermometers mounted at different locations near the PSB-a are in
progress, to attempt to identify any unique thermal feature that could
cause a slow glitch.

One of the bolometers (100-1a) shows another category of glitch that
is not detected in any other bolometer. These particular events have a
slow response, which is similar to the slow tail of short glitches,
but do not show a fast decaying part. They also have a rise
time constant of around 10\,ms. Such events are not accounted for in
our glitch subtraction method and this induces a small excess in the
residual power spectra at frequencies around 0.1\,Hz (see
Fig.~\ref{fig:spres}, and Sect.~\ref{methodres} for a discussion of
the impact of such events). This excess is accounted for in the total
noise budget for cosmological studies \citep{planck2013-p08}.

Finally, precursors of slow HCEs are events that are only seen in
coincidence and when associated with slow HCEs (see Sect.~\ref{showers}
for a description). They are not accounted for by the glitch subtraction
method, but are rare enough that they do not affect the cosmological
data analysis.

\section{Conclusions}

In this paper we have proposed an interpretation of the source of the glitch signals
observed in \Planck-HFI bolometer data, we have presented details of our
glitch subtraction and flagging method, and we have evaluated the impact of
residual glitches on cosmological data after processing.

We observed three families of glitches, ``short,'' ``long,'' and
``slow,'' where each is described by a sum of three or four
exponentials. They can be distinguished from each other by the
relative amplitudes and time constants of these exponential terms.
The rate of each glitch type is anticorrelated with solar activity,
indicating that they are all due to energy deposition by Galactic
cosmic rays. Using the response of the on-board particle monitor and
cryogenic stages during solar flares, we find that the spacecraft and
surrounding instrument mass shields the 0.1\,K plate and bolometers
from all particles with energies less than $39$\,MeV.

Short glitches have a response similar to the optical response of the
bolometers.  These short glitches result from direct particle energy
deposition in the grid, for the lowest energies, or in the Ge thermistor
for the highest energies.

Long glitches dominate the total glitch count rate, but are usually
smaller in amplitude than the short glitches.  Long glitches show a
fast part with about a 10 ms time constant, followed by the sum of
slow exponentials, with around 50\,ms to 2\,s time constants.  We find
that the flux of the long glitches is the same for all bolometers,
when we scale with the area of the Si die. Coincident counts in the
PSBs paired in a single bolometer module in the flight data, as well
as laboratory tests of spare bolometers on the ground using alpha
sources to expose the absorber grid and Si die separately, and tests
using a 23\,MeV proton beam line, support the conclusion that long
glitches are due to particles depositing energy in the Si die.  We
find a minimum energy deposition in the count distribution of long
glitches that corresponds to a proton with an energy at the minimum of
the proton energy loss spectrum traversing the Si die normal to the
surface. This minimum deposited energy is above the noise threshold
for about half of the detectors. Thus we can individually detect
almost all of the glitches in the bolometer data.  We suspect that the
fast part is due to hot ballistic carriers propagating from the wafer
to the thermistor. The slower timescale response is from the thermal
relaxation of the Si wafer.  We find evidence of long glitches caused
by Galactic cosmic ray He, but no evidence of Galactic cosmic ray
electrons. Low energy showers of secondary or delta electrons were
earlier suspected to be the cause of the high rate of glitches in
HFI. However, the high rate in PSBs is measured in coincidence between
the two bolometers. This signature cannot be caused by a delta
electron shower.  On the other hand, there {\it is\/} evidence of
delta electrons in the coincidence of short glitches in PSB pairs. In
this process, the delta electrons are produced when a high energy
particle hits the absorber grid of one bolometer and produces a
coincident glitch at high probability and at lower energy, in the
neighbouring grid.

Slow glitches are not understood at this time and have not been
replicated in the laboratory.  These glitches have nearly the same
long timescale behaviour as long glitches and are thus expected to be
due to energy deposition somewhere else in the module.

We detect high energy secondary showers from very high energy events,
both through coincident glitch detection among many bolometers and in
heating of the 0.1\,K plate. The rate (about $5\,{\rm h}^{-1}$) and
energy of these events is consistent with the impact of Galactic
cosmic rays at very high energies, $>$1\,GeV per nucleon.

We also detect a second type of high energy event with a different time
response than that of the secondary showers.  These have a characteristic
energy of around a TeV and a rate that is too high to be due to cosmic
rays. We currently do not have an explanation for the cause of these
events.

The new glitch analysis presented here is a considerable improvement
over the one previously reported. The method iterates between sky
signal estimated ring by ring independently, and glitch detection
above 3.2 times the rms noise after subtraction. The sky signal is
carefully estimated before subtraction using spline interpolation to
avoid spurious event detection. For a better separation of glitches
into different categories, we employ for each detected event a joint
fitting of the three glitch types, using templates. Long and slow
glitch templates are subtracted from the data. Without this
subtraction, residual glitch tails, after flagging would dominate over
the noise for a large range of timescales. The glitch removal method
improves the noise performance of HFI; we found that the residual
contribution from glitches after template subtraction is below the
noise at all frequencies, and has a maximum amplitude at around
0.1\,Hz. Realistic simulations, including the modelling of glitches,
support this result, and show for the two bolometers studied at
143\,GHz that residual glitches contribute to less than 20\% of the
noise spectrum on scales (along the scan) larger than about
$2^\circ$. At smaller scales the residual contribution is smaller than
5\% and is induced by long glitches below the threshold.  This
conclusion is supported by the observed correlations between the
signals from bolometers in a PSB pair.

The rapid part of all detected events is flagged. Consequently, the
amount of data rejected from the astrophysical analysis varies from
8\% to 20\%, depending on the bolometer.  This should be contrasted
with the ${>}\,95\%$ of acquired samples that are affected by glitches
in the unprocessed data.

With the help of the simulations, we have evaluated potential biases
introduced by our method on the sky signal estimation. We find no
evidence of bias in the cosmological signal and we estimate an upper
limit of $10^{-4}$ on the CMB anisotropy power spectrum.


\begin{acknowledgements}

  \Planck\ is a project of the European Space Agency -- ESA -- with
  instruments provided by two scientific Consortia funded by ESA
  member states (in particular the lead countries: France and Italy)
  with contributions from NASA (USA), and telescope reflectors
  provided in a collaboration between ESA and a scientific Consortium
  led and funded by Denmark.
  The development of \Planck\ has been supported by: ESA; CNES and
  CNRS/INSU-IN2P3-INP (France); ASI, CNR, and INAF (Italy); NASA and
  DoE (USA); STFC and UKSA (UK); CSIC, MICINN, JA and RES (Spain);
  Tekes, AoF and CSC (Finland); DLR and MPG (Germany); CSA (Canada);
  DTU Space (Denmark); SER/SSO (Switzerland); RCN (Norway); SFI
  (Ireland); FCT/MCTES (Portugal); and PRACE (EU). A description of
  the Planck Collaboration and a list of its members, including the
  technical or scientific activities in which they have been involved,
  can be found at
  \url{http://www.sciops.esa.int/index.php?project=planck&page=Planck_Collaboration}.

\end{acknowledgements}

\bibliographystyle{aa}

\bibliography{Planck_bib,P03e_extra}

\raggedright
\end{document}